\documentclass[10pt,journal,compsoc]{IEEEtran}
\usepackage{amsmath,amssymb,amsfonts}
\usepackage{algorithmic}
\usepackage{graphicx}
\usepackage{textcomp}
\usepackage{xcolor}

\usepackage{setspace}
\usepackage{algorithm}
\usepackage{tikz}
\usepackage{amsmath}
\usepackage{subcaption}

\usepackage{graphicx}
\usepackage{url}

\ifCLASSOPTIONcompsoc
  \usepackage[nocompress]{cite}
\else
  \usepackage{cite}
\fi
\ifCLASSINFOpdf
\else
\fi
\begin{document}
%
\title{\Large \bf {\sc DPCN:} Towards Deadline-aware Payment Channel Networks}
%
%
%
%

\author{Wenhui Wang,
        Ke Mu,
       Xuetao Wei 
\IEEEcompsocitemizethanks{\IEEEcompsocthanksitem Wenhui Wang, Ke Mu, Xuetao Wei are with the Department of Computer Science and Engineering, Southern University of Science and Technology, Shenzhen,
China, 518000.\protect\\
}
}

\IEEEtitleabstractindextext{%
\begin{abstract}
Payment channel is a class of techniques designed to solve the scalability problem of blockchain. By establishing channels off the blockchain to form payment channel networks (PCNs), users can make instant payments without interacting with the blockchain, avoiding the problems of long transaction consensus delays and high transaction fees. Recently, the optimization of PCNs has mainly focused on improving the network throughput via multi-path routing. However, the transaction's atomicity comes at a non-trivial cost for transaction completion latency that affects user experience in deadline-sensitive applications of PCNs. In this paper, we propose a new and systematic framework DPCN to consider the deadlines of transactions for payment channel networks while improving the success ratio of transactions. DPCN is enabled via a synergy of three components: (1) deadline-based dynamic transaction split mechanism that splits the transaction according to current network status and the transaction's deadline; (2) deadline-aware transaction scheduling that prioritizes near-deadline transactions; (3) deadline-aware transaction congestion avoidance algorithm, which uses a path window to balance transactions with different deadlines. Our extensive experiments show that compared with existing methods, DPCN can well meet the needs of transactions with different deadlines and ensure a higher success ratio for transactions in the payment channel networks.
\end{abstract}

}

\maketitle

\IEEEdisplaynontitleabstractindextext

%
\IEEEpeerreviewmaketitle

\IEEEraisesectionheading{\section{Introduction}\label{sec:introduction}}

%
%
%
%
\IEEEPARstart{D}{espite} many efforts to improve its efficiency \cite{rapidchain, pbft, monoxide, algorand}, blockchain still suffers from poor scalability, which introduces high latency to decentralized applications. For instance, the Bitcoin\cite{Bitcoin} network can only process 7 transactions per second on average and the Ethereum\cite{eth} network just provides 15 transactions per second. In contrast, Visa\cite{visa} can process 1700 tx/s on average and over 47000 tx/s at the peak. Blockchain is hard to scale since the underlying consensus protocol is not efficient enough. Transactions are passed through the network and need to be processed by all participants in the blockchain to maintain the consistent state change in a decentralized environment. 

At present, there are two major solutions to handle the scalability challenge in blockchain, namely, first-layer solutions (L1) and second-layer solutions (L2). First-layer solutions mainly focus on the optimization of the underlying design of blockchain, while the second-layer solutions aim to transfer parts of storage or computation tasks to other platforms to scale up the throughput of blockchain. A leading proposal among the second layer solutions is called the payment channel.
A payment channel is one of the cryptocurrency transactions that enables users to deposit or escrow their funds on the blockchain for payments with a prespecified user on any off-chain platform. For example, if Alice and Bob attempt to make transactions in a payment channel, they need to deposit some funds on the blockchain first to establish a channel. Then, when the channel is open, Alice and Bob can send signed transactions to each other through this channel without any verification on the blockchain.

To get the deposit funds back from the blockchain, Alice or Bob can send the most recent signed transactions to the blockchain to close the payment channel.
A network composed of multiple payment channels is called payment channel networks (PCNs). PCNs allow users that do not have direct payment channels in between to complete their transactions through a multi-hop path in the network. PCNs have been receiving growing attention these years and have also been deployed by many popular blockchains to improve the throughput and reduce the latency, for example, the Lighting network\cite{lightingnetwork} used for Bitcoin and the Raiden network\cite{raidennetwork} for Ethererum. 

Existing approaches\cite{flare,silentwhispers,SpeedyMurmurs,flash} mainly focused on the routing problem in PCNs to provide high-throughput transactions, e.g., multi-path routing by splitting the transaction into sub-transactions that are routed along multiple paths. However, the requirement of transaction atomicity comes at a non-trivial cost for latency, due to either network congestion or half-way incomplete sub-transactions. Such latency is unbearable for some deadline-sensitive applications (For example, instant transactions on exchanges\cite{coinbase,binance}, interactive blockchain games\cite{cryptoblades,Splinterlands}). Therefore, it is strongly desired for PCNs to allocate and schedule resources to meet the deadline requirements of these time-sensitive payment applications. Deadlines of transactions were considered in the single-hop optimal transaction scheduling problem\cite{scheDeadline}. Approaches \cite{d2tcp,DCTCP} proposed for the Internet cannot be simply applied to the PCNs, as the problem we solve in PCNs differs from it in crucial ways.  In the Internet, the capacity of each link is unchanged when sending packets. However, payment channels can only be able to send transactions by moving a finite amount of funds from one end to another in the channel. Thus, the capacity of each channel (the maximum funds it can send) varies dynamically after every payment in the channel. As a consequence, PCNs are more dynamic than the Internet and it is more challenging to guarantee transactions' completion time in PCNs.

In this paper, we present DPCN, a new and systematic framework that considers the deadlines of transactions in the payment channel networks while improving the success ratio of transactions. DPCN is enabled via a synergy of three components: (1) \textbf{deadline-based dynamic transaction split mechanism:} we find that the completion time of a transaction is affected by the size of the transaction being split into.  Deadline-based transaction split runs dynamically according to current network and transaction status. We use the window size of the path to correspond to the state of the path. We estimate the average processing time required for the path to process one transaction based on the completion of previous transactions on this path. Based on both the window size of the path and the ratio of the deadline of the transaction to the average processing time of the transaction on this path, the transactions of different deadlines are split into different sizes. Namely, the shorter the transaction's deadline is, the larger the 
amount of fund-units of split sub-transactions is; (2) \textbf{deadline-aware transaction scheduling:} deadline-aware transaction scheduling algorithm runs according to the remaining time of the transaction. We give priority to transactions with shorter deadlines. The transaction with the shortest deadline is picked up first from the waiting queue of the channel and sent to the channel for processing; (3) \textbf{deadline-aware transaction congestion avoidance algorithm:} when a transaction is queued at an intermediate node(a router) in the network for more than a certain period of time, we mark the transaction. This indicates that congestion is likely to occur in the network. This marked information will be forwarded to the sender together with the confirmation information after the receiver receives the transaction. For each path, we use the marked information and the completion status of the transaction in the most recent period of time to maintain a window (same as the window mentioned in deadline-based dynamic transaction split). We dynamically adjust the size of the window to limit the number of transactions sent to the path, so as to avoid the occurrence of congestion as much as possible, which ultimately avoids transaction failures due to long transaction processing time caused by network congestion. We conduct extensive experiments to evaluate DPCN under various settings, e.g., different topologies, channel sizes and transaction sizes. Experimental results show that DPCN is a very promising framework for deadline-aware payment channel networks.

Our contributions can be summarized as follows:
\begin{itemize}
    \item We propose DPCN, a new and systematic framework to consider the deadlines of transactions in payment channel networks for time-sensitive applications while improving the success ratio of transactions.
    \item The key novelty of DPCN is a synergy of three components, namely, deadline-based transaction split, deadline-aware transaction scheduling, and deadline-aware transaction congestion avoidance algorithm. Our framework DPCN synergizes these techniques to effectively improve the success ratio of transactions while meeting their deadline requirements.
    \item We prototype DPCN and extensive results show that DPCN can improve the success ratio of transactions compared to the state-of-the-art work.
\end{itemize}

\section{Background and Motivation}
\label{back}
\vspace{2mm}
\subsection{Payment Channel}
\begin{figure}[h]
  \centering
  \includegraphics[width=\linewidth]{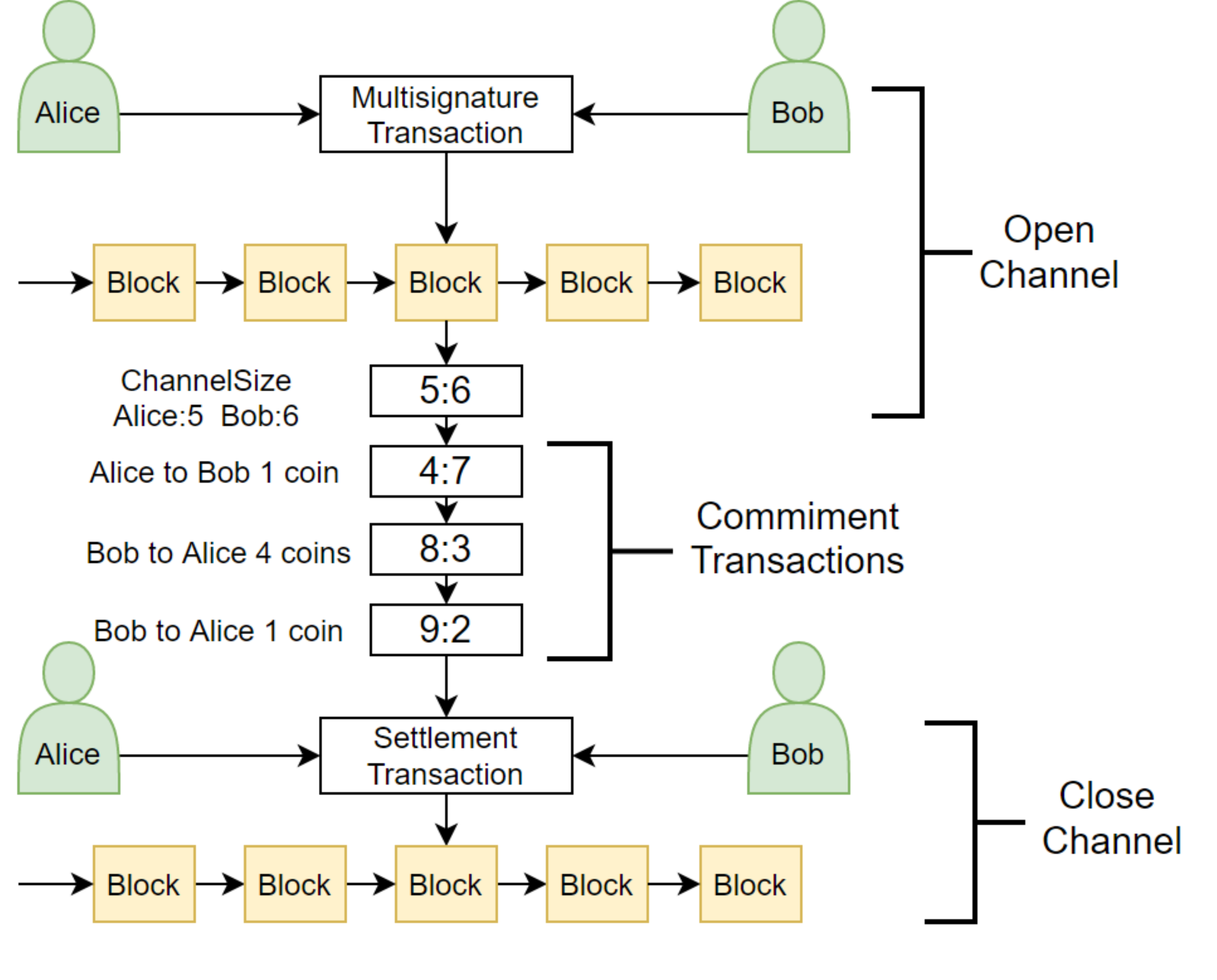}
  \caption{An example of a payment channel. \normalfont{First, Alice deposits 5 coins, Bob deposits 6 coins, and the initial state of the channel is (Alice:5, Bob: 6). Afterward, they make multiple transactions. Finally, the channel is closed, and the final status is updated to (Alice: 9, Bob: 2).}}
  \label{paymentchannel}
\end{figure}

\vspace{2mm}
The payment channel is one of the second layer solutions that move transaction processing away from the blockchain. A payment channel is established by two parties and allows them to send transactions to the blockchain only when the channel is opened and closed. After a channel is opened, users can conduct any number of transactions without participating in the consensus on the chain, which greatly improves the transaction throughput and reduces delays. The life cycle of a payment channel can be divided into three parts: channel establishment, transaction submission, and channel closure. Figure \ref{paymentchannel} shows an example of a simple bidirectional payment channel (a payment channel can be bidirectional or unidirectional). 
In order to establish a channel, two parties construct a smart contract on the blockchain and deposit a certain amount of funds into a multi-signature wallet. Then, they create a transaction. The approval of this transaction on the blockchain initiates the channel. 
After that, in the transaction submission phase, both parties can use this channel to send multiple payments without interacting with the blockchain. The transaction used in payment channel is called commitment transaction and needs to be signed by two parties. It updates the state of the channel like a transaction on the blockchain. Any party can submit a settlement transaction to the blockchain, or if one of the parties accidentally disconnects, the single party will submit the final commitment transaction to the chain. The settlement transaction represents the final state of the channel and is settled on the blockchain. 

\subsection{Payment Channel Networks}
\vspace{2mm}
\begin{figure}[h]
  \centering
  \includegraphics[width=\linewidth]{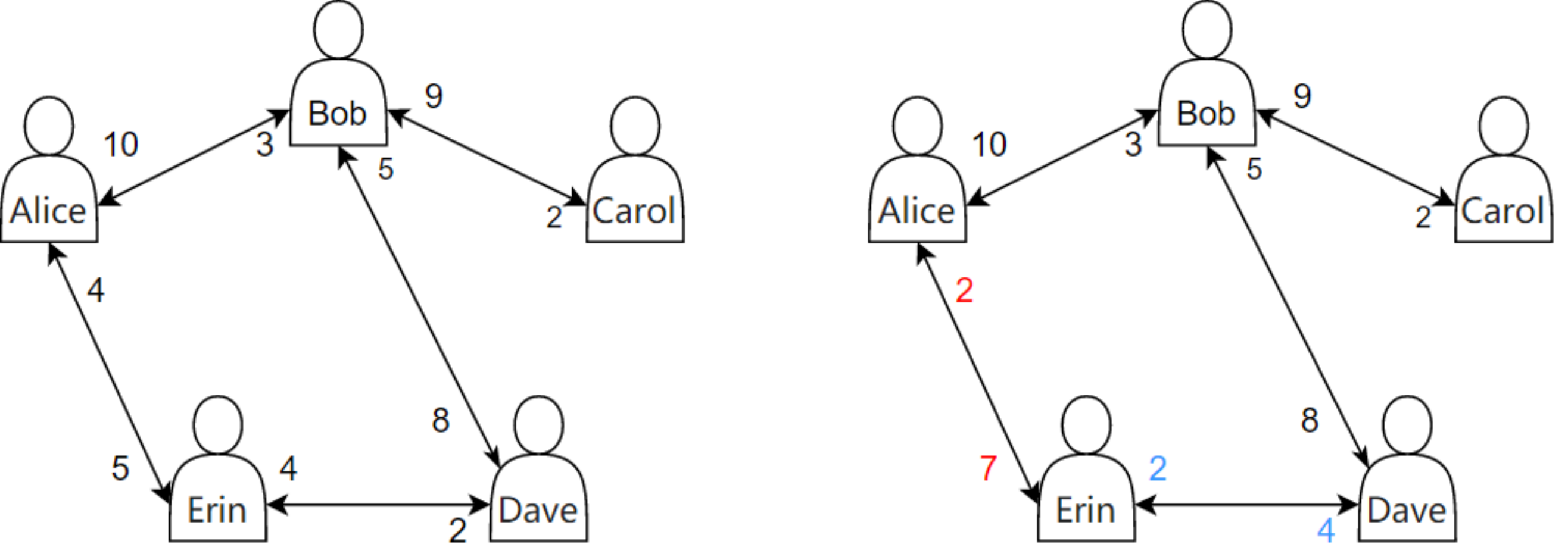}
  \caption{An example  of PCNs: Alice sends 2 coins to Dave through Erin, the left figure shows the initial state of PCNs, the right figure shows the state after the transaction is completed.}
  \label{paymentChanelNetwork}
\end{figure}

Since a certain amount of funds are required to be locked after the channel is established, and a long consensus process on the blockchain is required when creating and closing channels, it is impractical to establish payment channels between all pairs of parties who want to make a transaction. If two users want to conduct a transaction but do not have a directly connected payment channel, the transaction needs to be relayed by intermediate nodes to complete the multi-hop payment. In this case, multiple payment channels need to be used. A collection of bidirectional payment channels forms payment channel networks (PCNs). Figure \ref{paymentChanelNetwork} shows a simple payment channel networks. Suppose Alice wants to send coins to Dave multiple times. Since there is no directly connected channel between them, Alice needs to find a path to complete the transaction. There are two paths available in the current network: 1. Alice $\rightarrow$ Erin $\rightarrow$ Dave. 2. Alice $\rightarrow$ Bob $\rightarrow$ Dave. Taking path 1 as an example, Erin as an intermediate node, the entire transaction will be split into two parts: Alice sends 2 coins to Erin; Erin sends 2 coins to Dave. The hash-time-lock contract (HTLC)\cite{htlc} is usually used to ensure that the entire transaction can be completed, that is, to prevent Erin from not forwarding the transaction after receiving the coins, so as to ensure the security and the atomicity of the entire transaction is achieved. In the example above, the HTLC guarantees that if and only if after Dave receives the coins from Erin, Erin can get the coins sent from Alice. Otherwise, after a certain period time, the entire transfer process fails and Alice will be refunded.

\subsection{State of the Art: Spider}
\vspace{2mm}
Spider\cite{spider} considered packetizing transactions and multi-path routing transactions. It mainly includes two parts: (1) by setting the maximum transmission unit (MTU), Spider divides a transaction into $n$ parts regarding the size of the MTU, and routes these $n$ transactions through k paths to ensure that even the channel capacity is small, large transactions can still be completed, which greatly increases the success ratio of transactions and the throughput of the system; (2) by using the multi-path congestion control protocol to control the number of transactions sent to the PCNs, Spider marks transaction packets in the routing process and returns the marking information to the sender, thereby balancing the fund resources of each channel and avoiding channel unavailability due to exhaustion of funds.

\subsection{Motivation}
\vspace{2mm}


\begin{figure}
    \centering
  
    \includegraphics[width=\linewidth]{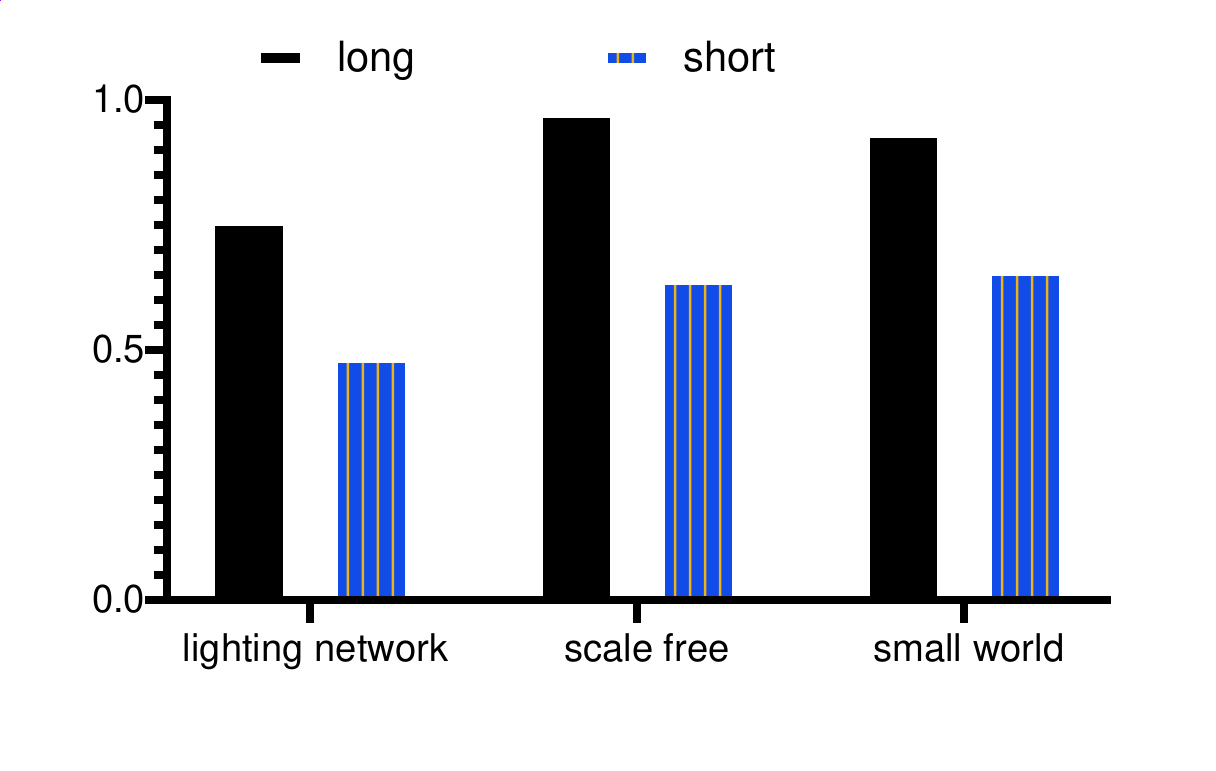}
    \caption{Success ratio for different topologies with the mean channel size 4000}
    \label{longShortRatio}
\end{figure}

\label{sec3}
Although Spider is successful in improving network throughput, it is a deadline-agnostic protocol, which does not distinguish between transactions with different deadlines. Without carefully considering such requirements, time-insensitive transactions are likely to be processed in priority over time-sensitive transactions. As a result, time-sensitive transactions may miss their deadlines in Spider, which ultimately impacts user experience.

We further present our motivation of the problem via experiments as shown in Figure~\ref{longShortRatio}. For the existing approach Spider, we choose three networks with different topologies, set the average channel capacity to 4000, and set different deadlines for transactions to conduct comparative experiments.  
In Figure \ref{longShortRatio}, we can see that, when the transaction deadline is longer, the system can obviously achieve a higher transaction success ratio. However, when we set a shorter deadline for some transactions, the transaction success ratio of PCNs drops significantly. That's because existing approaches are deadline-agnostic, the transactions with a longer deadline may be processed before those with a shorter one when PCNs are congested, causing the failure of time-sensitive transactions. Therefore, a deadline-aware framework is required to perform differentiated processing for transactions with different deadlines. This could ensure that near-deadline transactions can be completed as soon as possible. Actually, for a transaction, if one of the split sub-transactions fails, the entire transaction will fail. 

\section{ Design of DPCN}
\label{dpcn}
\vspace{2mm}
In this section, we propose a novel framework called deadline-aware payment channel networks DPCN. Focusing on transactions with different deadlines, we provide a well-designed mechanisms to achieve a higher transaction success ratio as well as minor impact on the throughput of payment network. In brief, DPCN is enabled via a synergy of three components: (a) deadline-based transaction split mechanism; (b) deadline-aware transaction scheduling; and (c) deadline-aware transaction congestion avoidance algorithm. 

\subsection{Prerequisites}

\vspace{2mm}
\noindent\textbf{Transaction Deadline:} Transaction deadline is the maximum time a transaction can be used from its initialization to completion in PCNs. If the processing time of transaction exceeds the deadline, the transaction will fail.

\vspace{2mm}
\noindent\textbf{Atomicity of Transactions:} In order to make better use of the resources of the network, we split a single transaction into multiple smaller sub-transactions and send them through different paths independently. As previous work\cite{flash,spider} did, we follow the assumption of the atomicity of the entire transaction can be guaranteed. It can be achieved by Atomic Multi-Path Payments (AMP) \cite{AMP}, the receiver won't be paid at all until all sub-transactions are completed. The design and implementation of such mechanism is beyond this paper.

\vspace{2mm}
\noindent\textbf{Transaction Confirmation and Cancellation: } For example, as shown in Figure \ref{paymentChanelNetwork}, for path Alice $\rightarrow$ Bob $\rightarrow$ Carol, if Alice wants to make a transaction $Tx$ with Carol through Bob. The transaction $Tx$ contains the transaction of $Alice$ and $Bob$ and the transaction of $Bob$ and $Carol$, respectively called $Tx_1$ and $Tx_2$. We say transaction $Tx$ has been confirmed only when both transaction $Tx_1$ and $Tx_2$ are completed, the fund-units in transaction $Tx$ is called the confirmed fund-units. Otherwise, these fund-units are called unconfirmed fund-units. Note that transaction $Tx$ is canceled when there is a failure in either transaction $Tx_1$ or $Tx_2$. Since the transaction has been cancelled, this part of the funds will be removed from the unconfirmed funds.

\vspace{2mm}
\noindent\textbf{Transaction Split Size: } If split size is $a$ and the fund-units of a transaction is $b$, a transaction is split into $\left \lfloor  a/b \right \rfloor $ sub-transactions of the fund-units $a$, and the remaining fund-units less than $a$ is used as a sub-transaction.

\begin{figure}[h]
  \centering
  
  \includegraphics[width=\linewidth]{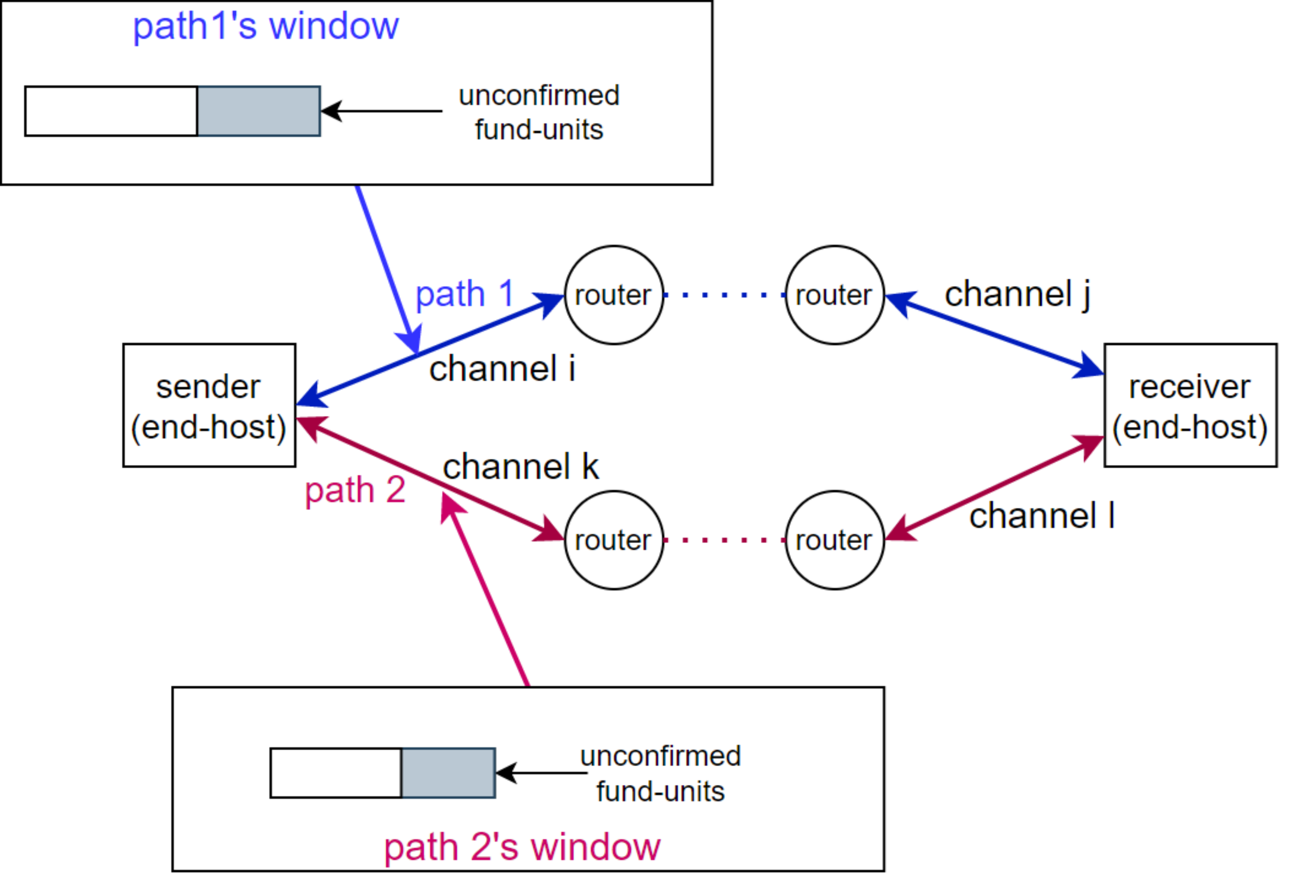}
  \caption{The windows corresponding to paths. There are two paths between the sender and the receiver and each path has a window}
  \label{smallWin}
\end{figure}

\vspace{2mm}
\noindent\textbf{Path Window Size:} We define path window size as the maximum fund-units that are not confirmed on a path (from sender to receiver) at any point of the time. 
Only when the sum of a transaction's fund-units and unconfirmed fund-units on the path is less than the window size, the transaction can be sent to this path. Otherwise, the transaction will enter into the queue of the sender and wait. As shown in Figure \ref{smallWin}, suppose there are $2$ paths between the sender and the receiver. For each path, we assign a window. Considering a transaction $Tx$ needs to send $f_1$ fund-units through path $1$. The unconfirmed fund-units on the path $1$ is $f_2$, and the window size on the path $1$ is $f_w$. Only when $f_1+f_2<f_w$, the transaction $Tx$ can be sent to path $1$ for processing. The window size can limit the transaction sending speed of the sender and avoid network congestion in the PCNs. Note that the window size is not the same as the remaining fund-units on a path, and must be smaller than the minimum remaining fund-units of any channel on this path.

\vspace{2mm}
\noindent\textbf{End Hosts:} End hosts are responsible for generating transactions according to the specified workload and sending the transactions to the PCNs. When an end host receives a transaction from other end hosts, it will return a confirmation message back along the same path. That is, all nodes of PCNs that wish to transact constitute end hosts, and each pair of end hosts acts as the sender and receiver of the transaction. We use the same prerequisite as previous work did\cite{flash,spider}: all end hosts have the knowledge of the topology and the capacity of each channel in PCNs, but do not know the current balance of the channels. End hosts compute appropriate paths based on the topology.

\vspace{2mm}
\noindent\textbf{Routers:} The routers, as intermediate nodes on the transaction path, are responsible for forwarding the transactions from the senders to the receivers and transferring the receivers' confirmation messages back to the senders. When a transaction is forwarded on a certain channel, the router locks the funds for it and decreases the corresponding channel capacity. The router only unlocks the transaction funds as it receives the confirmation message from final receivers, and then increases the channel's balance. If the channel corresponding to the router does not have enough funds to lock, the transaction will wait in the channel's queue until the transaction is selected for forwarding according to certain scheduling rules via our proposed deadline-aware transaction scheduling. When the queue is full, the arriving transaction will be dropped and a failure message will be returned to the end host.

\subsection{Deadline-based Dynamic Transaction Split Mechanism} 
\begin{figure}
    \includegraphics[width=0.85\linewidth]{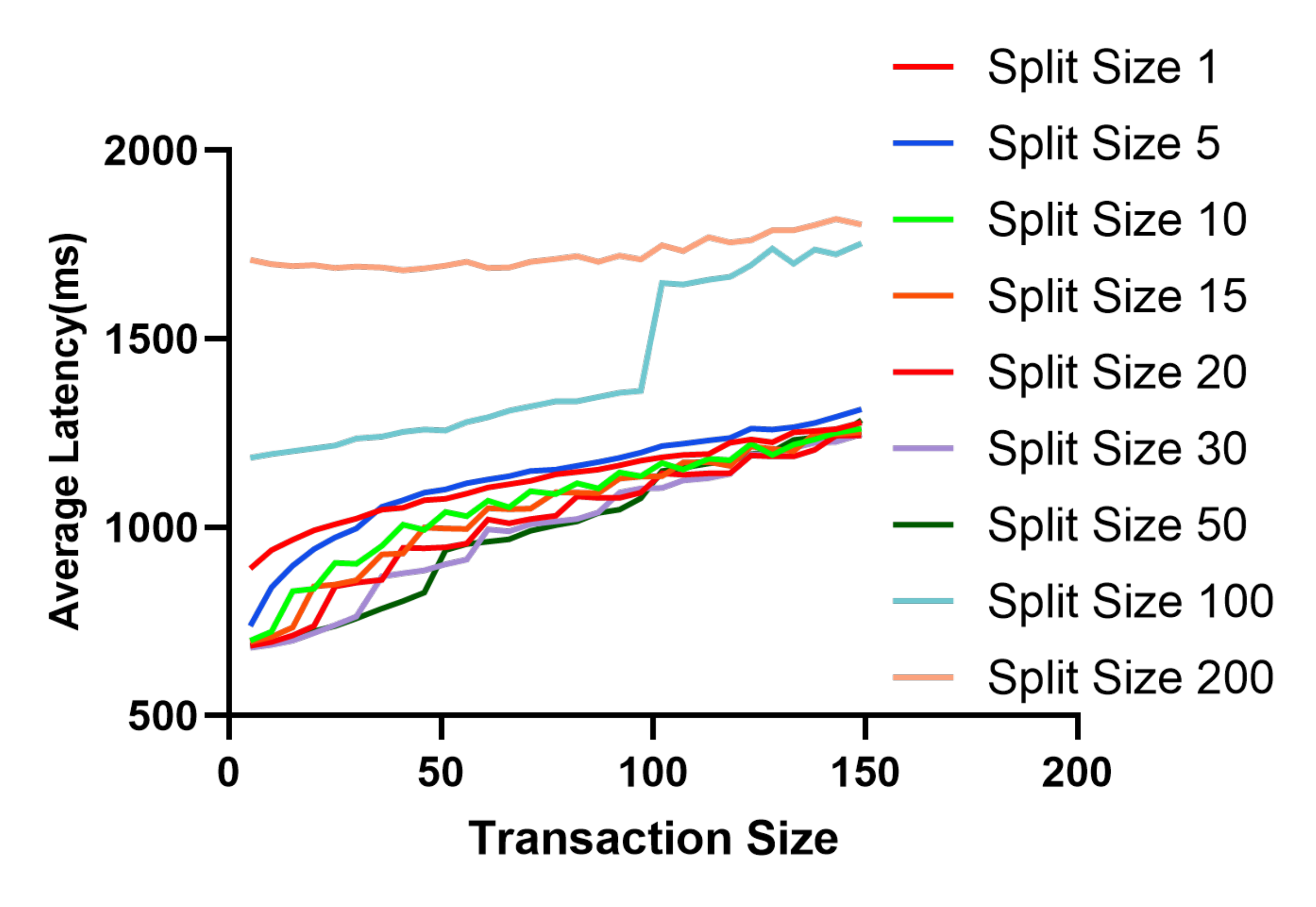}
    \caption{Average latency with different split sizes}
    \label{latencySplitSize}
\end{figure}

\begin{figure}
    \includegraphics[width=0.85\linewidth]{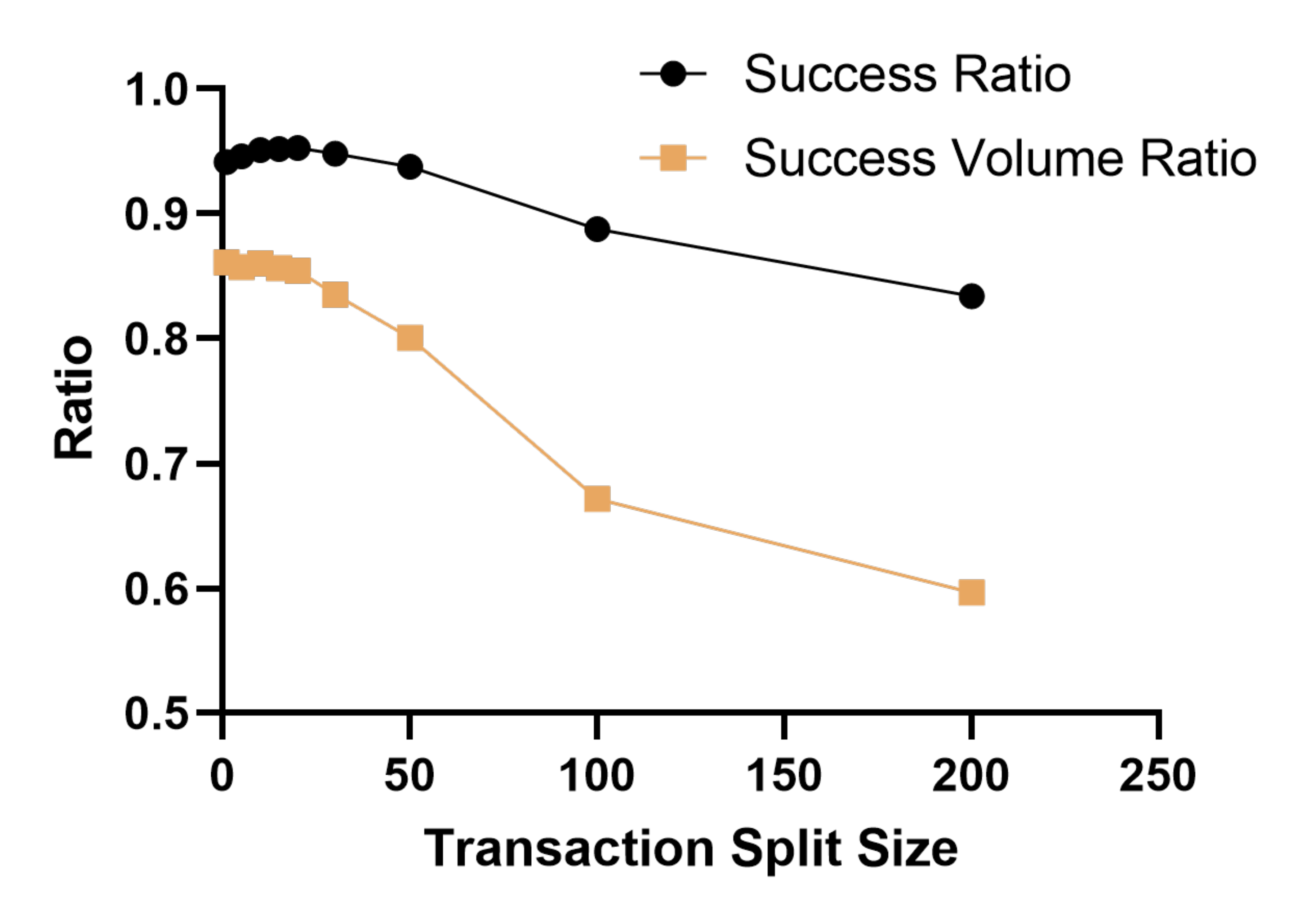}
    \caption{Success ratio with different split sizes}
    \label{successRatioSplitSize}
\end{figure}

The main challenge of the current transaction processing mechanism in PCNs is to find channels with sufficient balance on nodes to forward an incoming transaction. However, since the sender of the transaction does not know the current available balance of each channel, one direction of the channel may not have enough balance to forward the transaction, which will cause the transaction to fail. Therefore, in order to improve the success ratio of transactions and make better use of resources of the network, people try to split the transaction into multiple smaller sub-transactions and send them to different paths for processing. Previous work\cite{spider,waterfilling} has proved that this greatly increases the probability of successful transactions and increases the throughput of PCNs. We consider that different transaction split sizes may affect the completion time of the transaction. Intuitively, the greater the number of sub-transactions that the transaction is split into, the more time it needs to process the transaction, and the transaction delay may increase. To prove this hypothesis, we use the simulator in \cite{spider}, the topology(with 106 nodes and 265 payment channels) sampled from the Lightning Network and set the mean channel size to 4000 to test the average completion time of transactions of different sizes and the success ratio of all transactions for different transaction split sizes. We find that, although the split size of large transactions has little effect on the completion time, appropriately increasing the split size of the transaction can effectively reduce the completion time of the transaction for small transactions. As shown in Figure \ref{latencySplitSize} and \ref{successRatioSplitSize}, when the split size is $50$, compared to split size $1$, the average completion time of small transactions is reduced by more than $20\%$, while the successful volume ratio of transactions only drops by $7\%$. This is because, when splitting transactions into an excessive number of smaller ones in a network with large transactions handling capability, the growing transactions will cause delays to be prominent. However, when the transaction split size is too large, the success ratio of the transaction drops significantly. This is because, if the split size is too large, it is likely to exceed the available capacity limit of some channels in the network, resulting in transaction failure. Therefore, for transactions with different needs, an appropriate transaction split size should be selected according to the current network status.



Based on the experiments above, we observe that a reasonable transaction split size can effectively affect the transaction processing time. 
For instance, increasing the split size of a time-sensitive transaction appropriately can definitely reduce the number of transactions to be processed in the network, which can also reduce the transaction processing delay and increase the probability of transaction success. Instead of splitting them to the same size as prior work\cite{spider}, we dynamically split the size of the transaction.


We use a heuristic algorithm to split the transaction based on the three following conditions: the deadline of the transaction, the average completion time  and the average window size  of all paths associated with the transaction.

For a transaction $Tx$, the split size is as follows:
\begin{align}    \label{splitSize}
  SplitSize&=WS_{avg} \times \gamma \\
  \gamma &= \frac{1}{1 + c ^{- (x - b)}}  \label{split}\\
  x&=\frac{Tx_{deadline}}{P_{avgTime}} \label{x}
\end{align}

$WS_{avg}$ represents the average value of the window size of all paths between sender $Tx_{s}$ and receiver $Tx_{r}$. 
$P_{avgTime}$ represents the average value of the time required to complete one transaction on all paths. This time includes the entire time from the transactions were sent to completion, including the time when the transactions are being routed, waiting in the queue, and so on. $c$ and $b$ are two constants, which are used to control the size of the split.

\begin{figure}[h]
  \centering
  
  \includegraphics[width=0.8\linewidth]{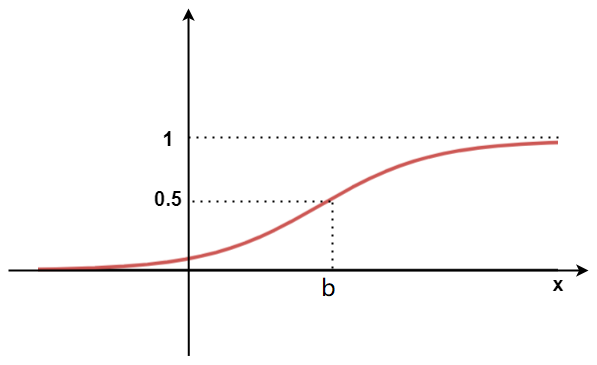}
  \caption{The function graph of $\gamma$}
  \label{splitSizeFunc}
\end{figure}

The window size of the path can well reflect the processing capacity of the path. Assuming that the sender of the current transaction $Tx$ is $Tx_{sender}$ and the receiver is $Tx_{receiver}$, we select all paths whose starting point is $Tx_{sender}$ and ending point $Tx_{receiver}$, and calculate the average window size $WS_{avg}$ and the average completion time $P_{avgTime}$ for a transaction (Note that, the calculation approach of the average completion time of a transaction on a path is given in \ref{txCongAvoAlg}.) We take $WS_{avg}$ as the upper limit of the size of a transaction split. After that, we calculate the split size of the transaction according to the deadline of the current transaction.

As shown in Equation \ref{x}, we calculate the urgency factor $x$ of the current transaction, and the larger $x$ indicates the current transaction is more urgent. Intuitively, when the $x$ is larger, we hope the transaction can be processed at a faster speed. Therefore, it needs to be split into multiple sub-transactions with larger fund-units that are smaller than is $WS_{avg}$. Conversely, a longer processing time has less impact on whether the transaction with a larger $x$ can be successfully completed. So it can be split into multiple sub-transactions with smaller fund-units. We define $\gamma$ as the split factor of the transaction, and the split size is $WS_{avg} \times \gamma$. The definition of $\gamma$ is shown in Equation \ref{split} and the functional image of $\gamma$ is shown in Figure \ref{splitSizeFunc}. The change trend of this function can well meet our requirements. As $x$ increases, the value of $\gamma$ gradually increases, that is, the size of the split gradually increases. Before $x$ reaches $b$, the growth rate gradually increases, and the growth rate reaches the maximum point when $x$ is equal to $b$. After that, the growth rate decreases. We consider that the more urgent the transaction is, the size of the transaction split is as large as possible, that is, we make the size of $\gamma$ grow as fast as possible. However, when the value of $x$ is too large, the transaction may not be completed eventually. Although we increase the size of the split, as $x$ increases, the growth of the split size, that is, the growth rate of $\gamma$ will be reduced. 
This will prevent transactions with a high probability of failure from occupying network resources too much.

\subsection{Deadline-aware Transaction Scheduling}
\vspace{2mm}
When the arriving transactions exceed the channel's processing capacity, they will wait in the channel's queue. Deadline-aware transaction scheduling aims to solve the issue: how to determine the processing order of queued transactions when the channel is congested?
In our framework DPCN, we mainly consider two related parameters to sort the transactions: remaining time ($RT$) and transaction size ($TS$). The remaining time represents the maximum time that the transaction can be processed in the PCNs from the current time.
Similar to the earliest deadline first\cite{EDF}, we firstly prioritize $RT$ as the first indicator, which means that the transactions with shorter remaining time will get higher processing priority. Then, when there is a tie or no deadline, we consider the transaction size in the following. We use $TS$ as the second indicator, transactions with smaller $TS$ will get higher priority in our approach. 



\subsection{Deadline-aware Transaction Congestion Avoidance Algorithm}
\label{txCongAvoAlg}
\vspace{2mm}
In order to alleviate network congestion caused by the excessive number of transactions in PCNs, we propose a congestion avoidance algorithm inspired from \cite{d2tcp,tcpip,tcpcc,icc} for PCNs that provides non-equivalent resources for transactions of different deadlines.
In our approach, when the waiting time of a transaction in the router of the PCNs exceeds a certain threshold $\delta$, the router will mark the transaction. The marked information will be forwarded to the sender together with the confirmation information after the receiver receives the transaction. This mark information indicates that the path in the network may be congested. Then, the sender can adjust the path's window size based on this feedback information to control the number of transactions sent to the path, so as to avoid network congestion, like the explicit congestion notification(ECN)\cite{tcpAndECN,ecn}.
The window size of each path represents their transaction processing capacity in the current state. It also shows the upper bound of the transaction amount that can be sent without confirmation on a certain path. 

Next, our problem is how to adjust the size of the window based on transaction's deadline. We use the following algorithm to adjust the window size of the paths connected to the end hosts to limit the number of transactions sent to PCNs.

For path $j$ in PCNs, we use $k$ to represent the proportion of marked transactions in the feedback information received by end hosts within a certain period $t_{interval}$. $e$ is a constant, which is the weight given to new sample $k$. We use $\alpha$ to represent the current congestion state of transactions in the network. The new congestion state  can be updated instantly by previous state $\alpha$ and current proportion of marked transaction $k$, as shown in Equation (\ref{eq:alpha}).

\begin{equation}
    \alpha = (1-e)\times \alpha + e \times k 
    \label{eq:alpha}
\end{equation}

Then, we try to estimate the time required to complete the transaction on the path $j$, which is represented by $ET_j$. Note that, unlike traditional computer networks, the completion time of a transaction on a path has nothing to do with the size of the transaction. This is because, as long as the path has enough balance, a large transaction and a small transaction are processed in the same way. When a transaction $Tx$ is completed on the path $j$, we calculate its completion time $t$ on this path, $t=Tx_{timeCompletion}-Tx_{timeSent}$. Based on the previous state of $ET_j$ and current time $t$, we update the value of $ET_{j}$, $y$ is a constant, which is the weight given to the new sample $t$.

\begin{equation}
    ET_{j} = (1-y) \times ET_j + y \times t
    \label{estimateTime}
\end{equation}

After obtaining $ET_j$, we can estimate the time required for a transaction to complete on the path $j$. We use $d$ to indicate the urgency of the transaction $Tx$. The larger $d$ is, the closer the deadline is. As is shown in Equation \ref{d}, if a transaction's estimated finish time is less than the deadline, the value of $d$ is less than $1$, otherwise, the value of $d$ is greater than $1$.
\begin{equation}
    d = \frac{ET_j }{Tx_{deadline}}
    \label{d}
\end{equation}
Finally, in order to change the size of the window, we define a penalty function $p$ related to $d$, as shown in Equation (\ref{p=ad}). Intuitively, $p$ increases the processing opportunity of transactions with shorter remaining time and decreases those with longer remaining time.
This function was originally proposed in \cite{dv}, which was called gamma-correction.

\begin{equation}
    p = {\alpha} ^ {d}
    \label{p=ad}
\end{equation}

Based on the penalty function $p$, we adjust the size of the window of the path $j$ based on the equation (\ref{Wi}). This equation is similar to the control mechanism of AIMD\cite{aimd} in TCP. Here $g$ and $\beta$ are constants, used to adjust the window size. Different from the ddl-agnostic window adjustment method in spider\cite{spider}, our method better matches the characteristics of transaction deadlines and thus can better guarantee the success rate of deadline-sensitive transactions.

\begin{equation}
    \begin{aligned}
    W&=\left\{
    \begin{array}{lll}
	W\times (1-\frac{p}{g}),& p > 0 ,\\
	W + \beta,& p = 0.\\
    \end{array} \right.
    \end{aligned}
    \label{Wi}
\end{equation}

In order to ensure that transactions can be completed on time under congestion, the number of transactions in PCNs should be reduced regarding their deadlines. Specifically, the number should be reduced to a greater degree if the transaction deadline is longer.

\begin{algorithm}[ht]
\setstretch{1.35}
\caption{Path window adjustment algorithm for end hosts}
\label{alg}

\begin{algorithmic}[1]
\REQUIRE ~~\\
 ack message: $aMsg$ \\
 totalTransactions: $totTx$ \\
 totalTransactionsMarked: $totTxMark$ \\
\ENSURE ~~
 Window size for $path_j$ $W$

\STATE $totTx \gets totTx + 1$
\IF{aMsg.isMarked} 
    \STATE $totTxMark \gets totTxMark + 1$
\ENDIF
\IF{$totTx > txThreshold$}
    \STATE $ k \gets \frac{totTxMark}{totTx}$
    \STATE $\alpha \gets (1-e)\times \alpha + e \times k$
    \STATE $totTx \gets 0$, $totTxMark \gets 0$
\ENDIF
\STATE $t \gets nowTime - aMsg.timeSent$
\STATE $estimatedTime \gets (1-y) \times estimatedTime + y \times t$ \\

\STATE $d \gets \frac{estimatedTime}{aMsg.deadline}$
\STATE $p \gets {\alpha} ^ {d}$
\IF{$p > 0$}
    \STATE $W \gets W\times (1-\frac{p}{g})$
\ELSE
    \STATE $W \gets W + \beta$
\ENDIF
\end{algorithmic}
\end{algorithm}
Note that $\alpha$ is a number less than or equal to $1$.  When the proportion of congested marked transactions increases (as $\alpha$ gradually increases to $1$), for transactions with a long remaining time (corresponding to $d<1$), the value of the penalty function increases rapidly. In addition, the window size is reduced, and the number of such transactions is also reduced rapidly. At the same time, for transactions with a short remaining time (corresponding to $d>1$), the value of the penalty function increases slowly, so as to ensure that this type of transaction can be sent as soon as possible. As shown in Equation \ref{Wi}, when there is congestion, the window will be reduced by $\frac{p}{g}$. It ensures that transactions with shorter deadlines have fewer backoffs than higher deadlines. When there is no congestion, the proportion of transactions with a congestion mark is $0$ ($\alpha=0$), the window is increased by $\beta$, which means that more transactions can be sent. 

The adjustment process of the window is as shown in Algorithm \ref{alg}. Data fields in $aMsg$ in algorithm\ref{alg} are shown in Table \ref{aMsg}.

\begin{table}[]
\caption{Data in the confirm message forwarded by routers}
\label{aMsg}

\resizebox{\linewidth}{!}{
\begin{tabular}{cclllll}
\hline
Field       & \multicolumn{6}{c}{Description}                                        \\ [5px]\hline
txTD        & \multicolumn{6}{c}{The split transaction ID}                           \\[5px]
receiver    & \multicolumn{6}{c}{The receiver of the fund-units}                           \\[5px]
pathIndex   & \multicolumn{6}{c}{The pathIndex for this transaction processing path} \\[5px]
timeSent    & \multicolumn{6}{c}{The time the transaction was sent}                  \\[5px]
isSuccess   & \multicolumn{6}{c}{If the transaction has been successfully completed} \\[5px]
amount      & \multicolumn{6}{c}{The number of fund-units for this transaction}      \\[5px]
hasDeadline & \multicolumn{6}{c}{If the transaction has deadline}                    \\[5px]
deadline    & \multicolumn{6}{c}{The transaction's deadline}                         \\[5px]
isMarked    & \multicolumn{6}{c}{If the transaction has been marked}                 \\[5px]
largerTxID  & \multicolumn{6}{c}{The ID of the transaction before the split}         \\ \hline
\end{tabular}
}
\end{table}

\section{Evaluation}
\label{eva}
In this section, we evaluate the performance of DPCN, and compare it with the state-of-the-art PCNs transaction processing approaches under different conditions. As mentioned in previous work \cite{spider}, IO bottleneck problem in Lighting Network Daemon(LND)\cite{lnd} causes variations of tens of seconds in transaction latency even on small topologies, which shows it cannot accurately test the system's performance. Therefore, we build our simulator based on the code in \cite{spider}, which has only a small difference in the success ratio from that in the real network.

\subsection{Experimental Setup}
Our experiments are conducted on a cluster of servers with two Intel® Xeon® Silver 4210 CPUs and 128GB memory. We build DPCN based on the code in \cite{spider}, which uses the OMNET++ simulator\cite{omnetpp} to model PCNs.

\vspace{2mm}

\noindent\textbf{Topology.} We use three topologies: Watts-Strogatz small world topology \cite{WattsStrogatz,sw} (50 nodes and 200 edges), scale-free Barabasi-Albert topology \cite{BarabasiAlbert}(50 nodes and 336 edges) and subnetwork(106 nodes and 265 edges) sampled\cite{graphSampled} from Lighting Network(LN) on July 15,2019(over 5000 nodes and 34000 edges). These topologies are the same ones used by \cite{spider}. For each channel, the capacity is sampled according to the actual capacity in the Lightning Network\cite{lightingnetwork} on July 15, 2019, and scaled up appropriately to generate a network structure with different average channel capacities. The values of average channel capacities are 900, 1350, 2750, 4000, and 8750, respectively. For the LN subnetwork, we set the transmission delay between 0.29ms and 130ms in the LN. In the other two networks, we set the transmission delay to 30ms for each link. 

\begin{figure}[h]
  \centering
 
  \includegraphics[width=0.85\linewidth]{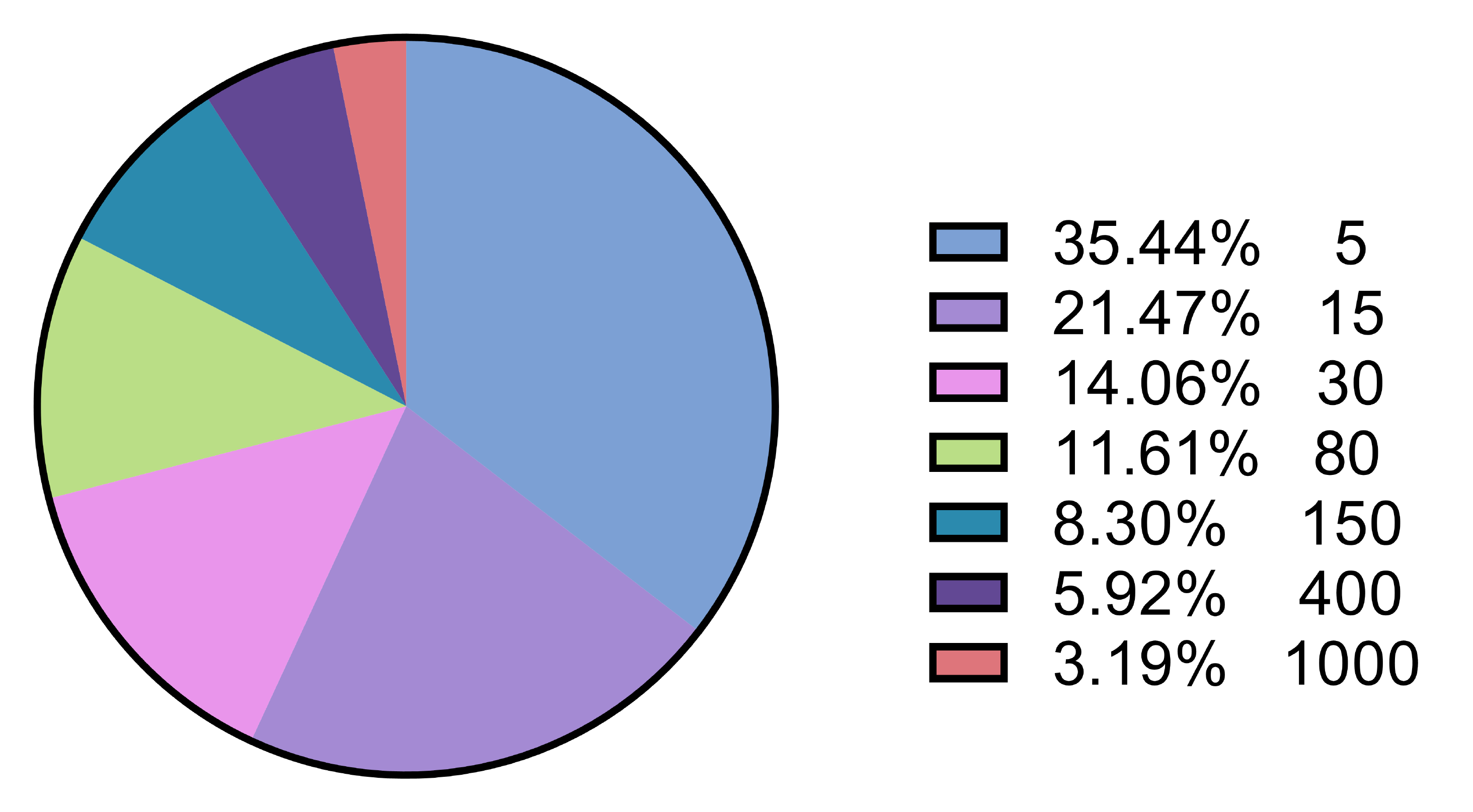}
  \caption{Transaction size  distribution}
  \label{workload}
\end{figure}

\vspace{2mm}
\noindent\textbf{Workload.} We set each end-host to send an average of 30 transactions per second, including 10 receivers, and transactions arrive as a Poisson process. For each set of experiments, we use five random different sets of transaction data and use the average as the result. Previous work\cite{flash,spider} pointed out that the distribution of transactions on the existing PCNs is highly skewed, namely, most of the transaction amount is small and the proportion of transactions with large amounts is small. Therefore, we use the proportion shown in Figure \ref{workload} to allocate the size of the transaction amount in the workload. If the transaction has a deadline, we set the deadline of the transaction as shown in Table \ref{TxDDL}. Note that, due to the lack of data in real scenarios, the deadlines of transactions in Table \ref{TxDDL} may be different from the real situation, but they can still reflect the characteristics of different transactions.

\begin{table}[h]
\centering
\caption{\label{TxDDL} Transaction deadline for different sizes}
\begin{tabular}{|c|c|c|c|c|c|c|c|}
\hline
 Size &5   & 15  & 30  & 80  & 150 & 400 & 1000 \\ \hline
 Deadline &0.6 & 0.7 & 0.8 & 0.9 & 1.1 & 1.5 & 2.0  \\ \hline
\end{tabular}
\end{table}

\begin{figure*}[ht]
    \centering
    
    \begin{subfigure}[b]{0.33\linewidth}
       
        \includegraphics[width=\linewidth]{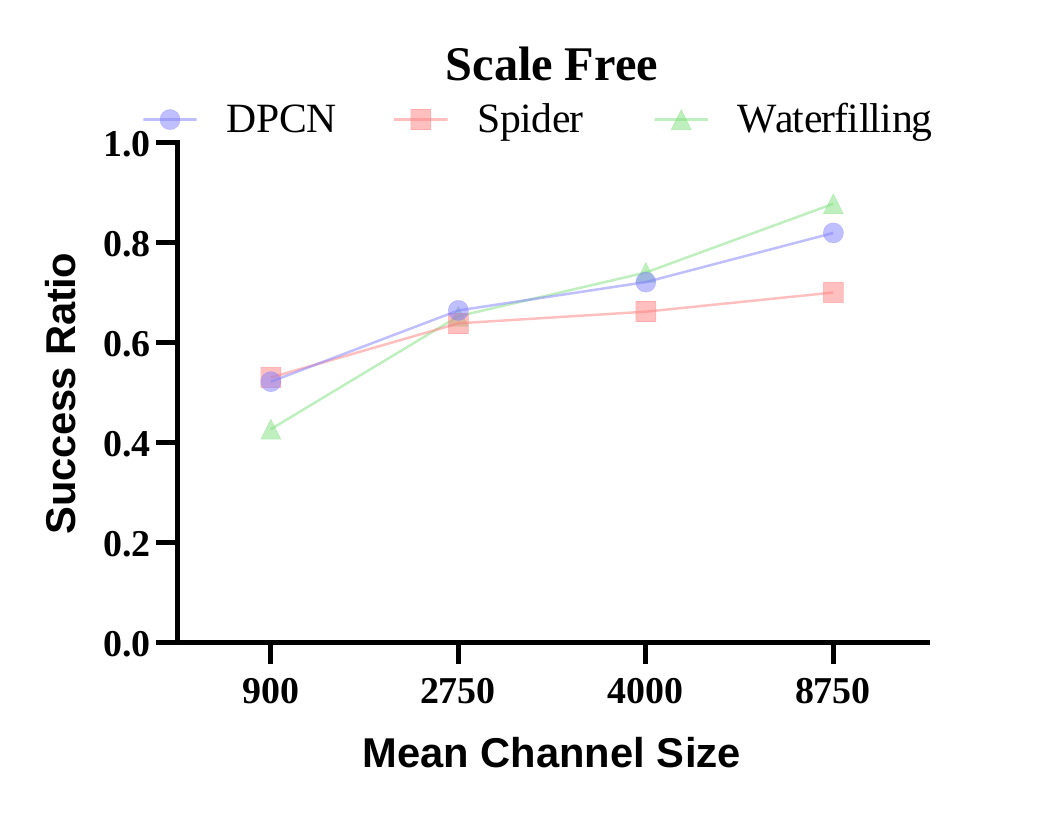}
        
        \label{sucRSF}
    \end{subfigure}
    \hfill
  \begin{subfigure}[b]{0.33\linewidth}
       
        \includegraphics[width=\linewidth]{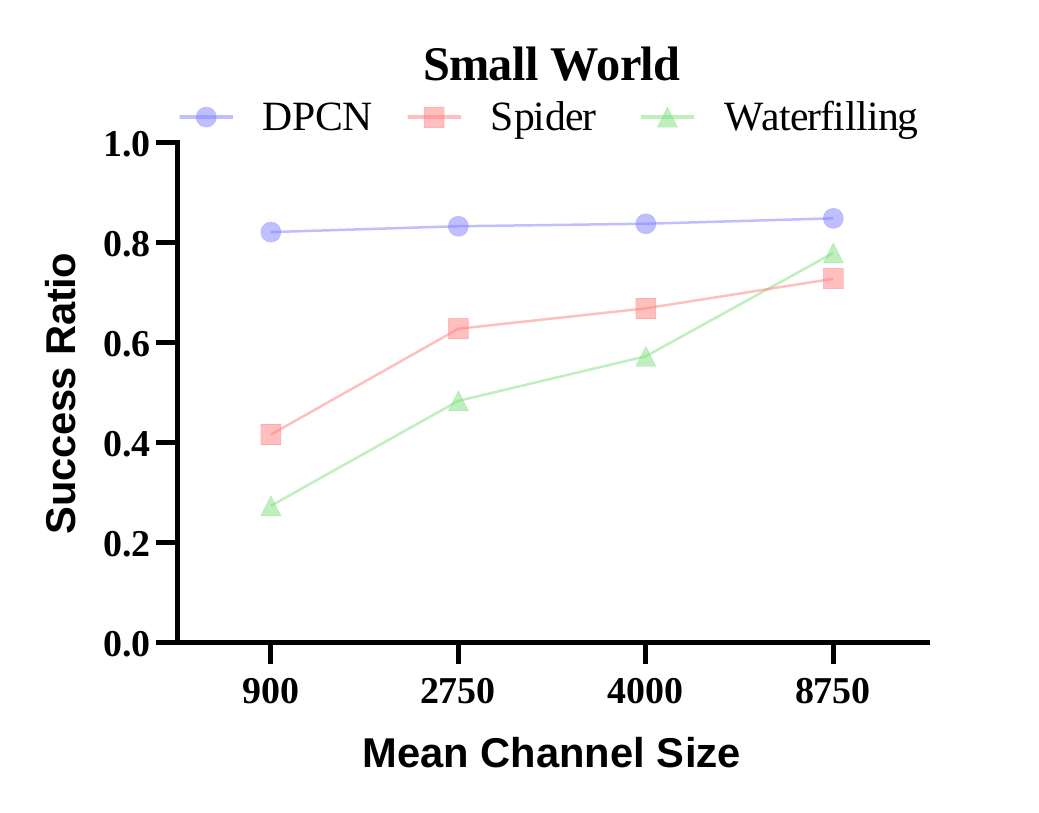}
        
        \label{sucRSW}
    \end{subfigure}
    \hfill
     \begin{subfigure}[b]{0.33\linewidth}
       
        \includegraphics[width=\linewidth]{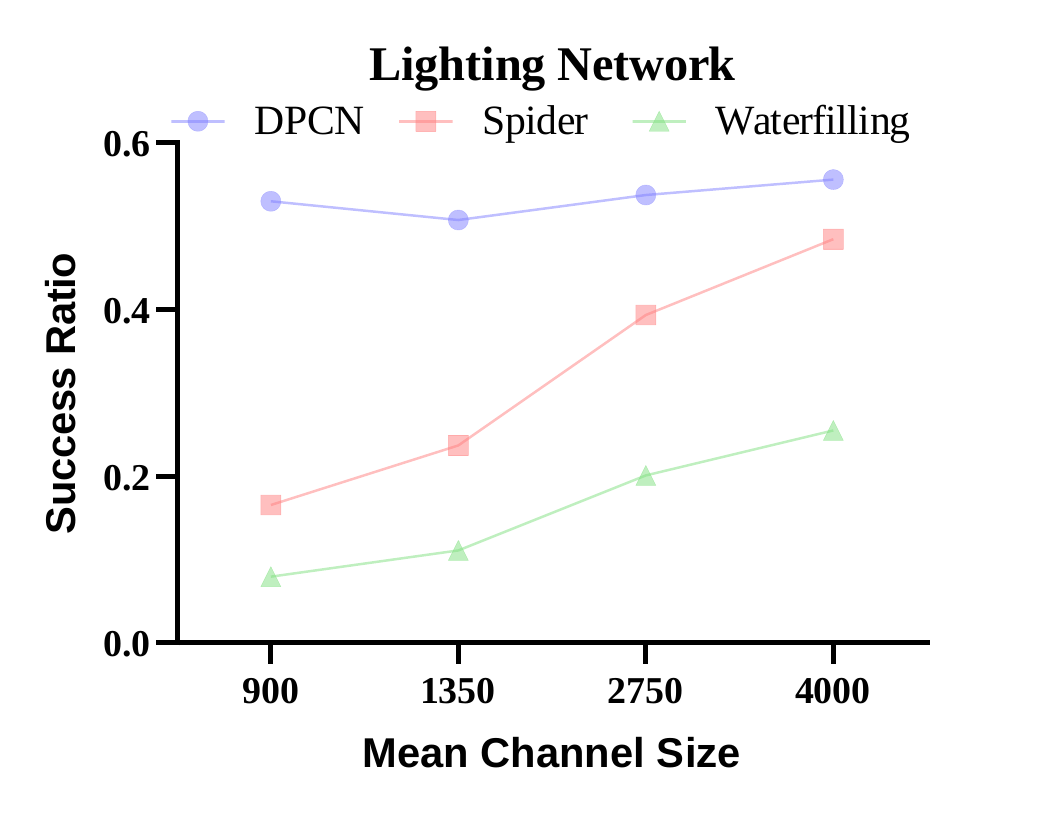}
        
        \label{sucRLND}
    \end{subfigure}
    \caption{Success ratio for three topologies}
    \label{sucR}
\end{figure*}

\begin{figure*}[ht]
    \centering
    
    \begin{subfigure}[b]{0.33\linewidth}
        
        \includegraphics[width=\linewidth]{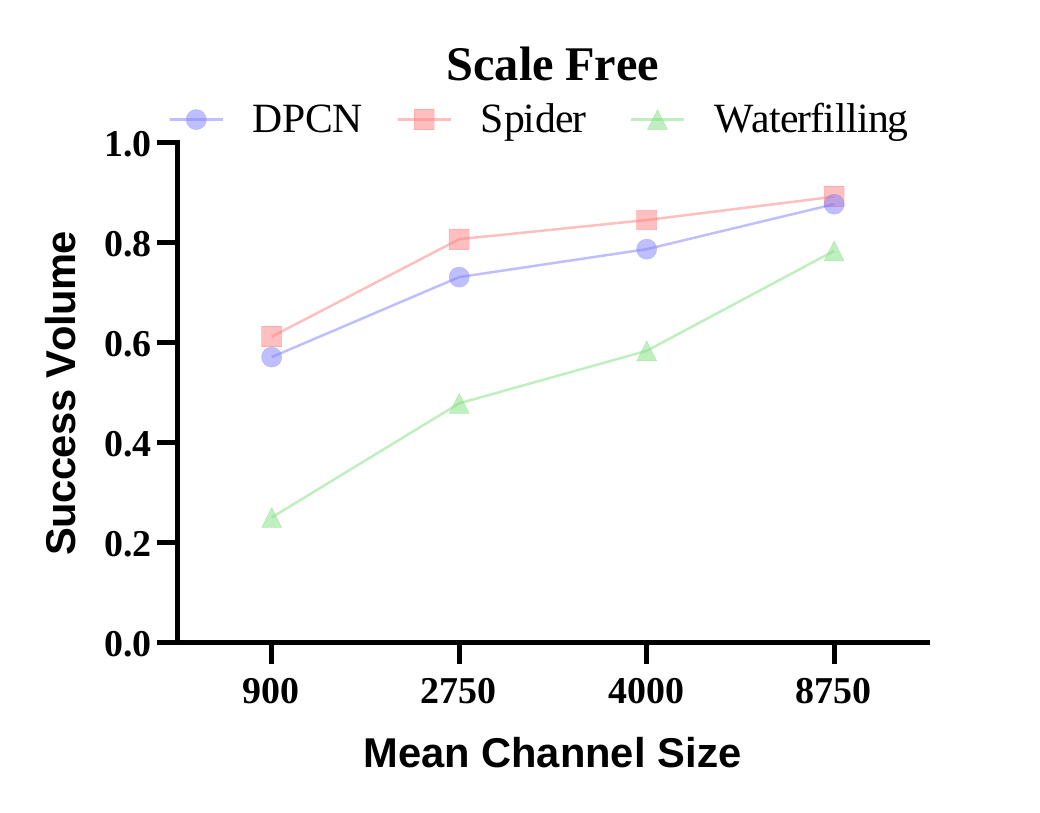}
        
        \label{sucVSF}
    \end{subfigure}
    \hfill
    \begin{subfigure}[b]{0.33\linewidth}
        
        \includegraphics[width=\linewidth]{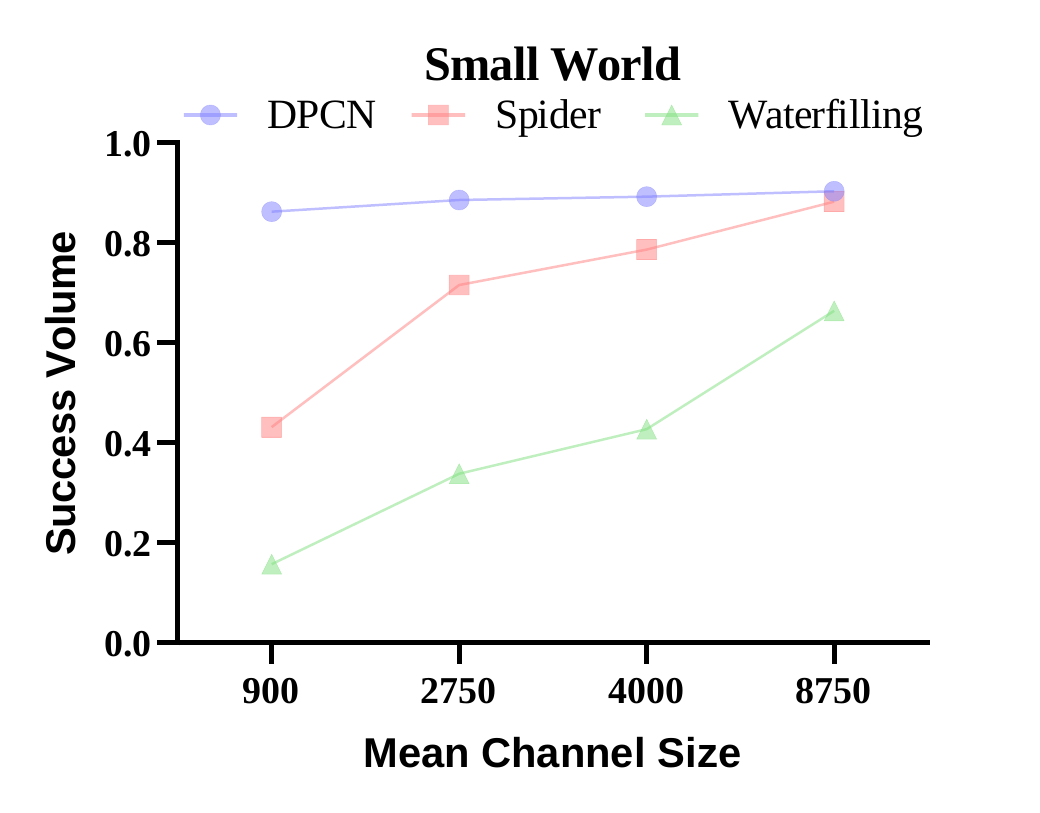}
        
        \label{sucVSW}
    \end{subfigure}
    \hfill
    \begin{subfigure}[b]{0.33\linewidth}
       
        \includegraphics[width=\linewidth]{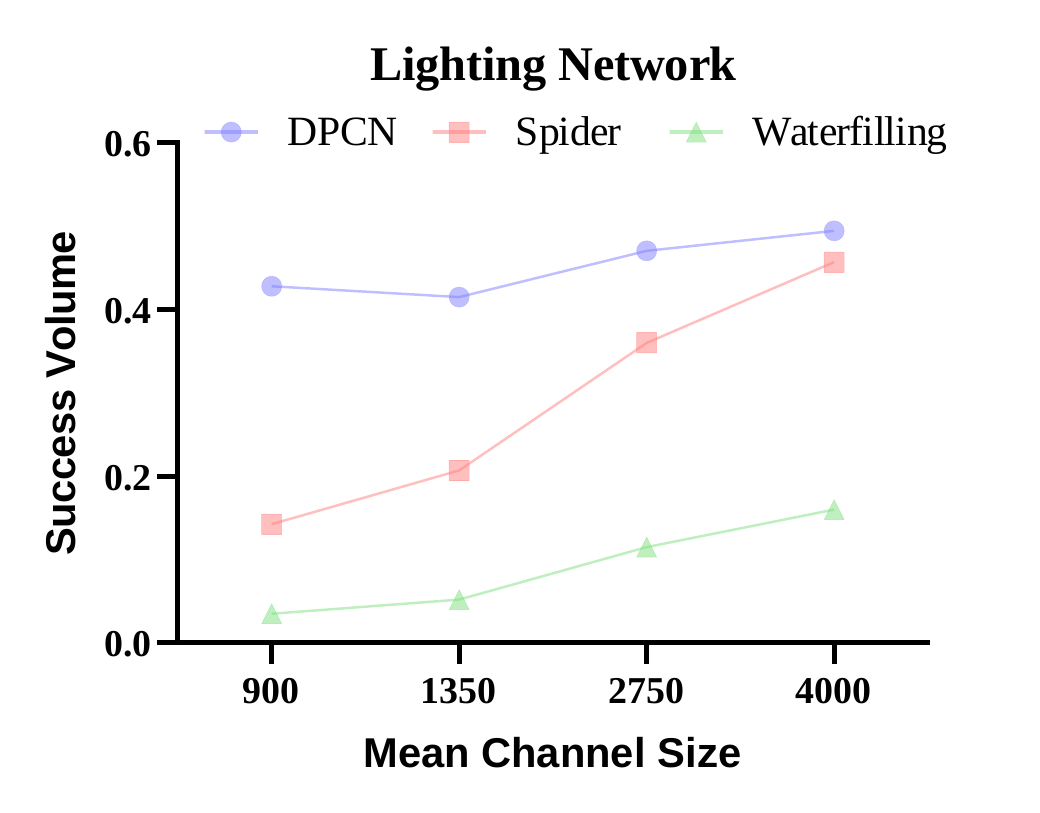}
        
        \label{sucVLND}
   \end{subfigure}
   \caption{Success volume for three topologies}
   \label{sucV}
\end{figure*}

\begin{figure*}[ht]
    \centering
   
    \begin{subfigure}{0.33\linewidth}
        \includegraphics[width=\linewidth]{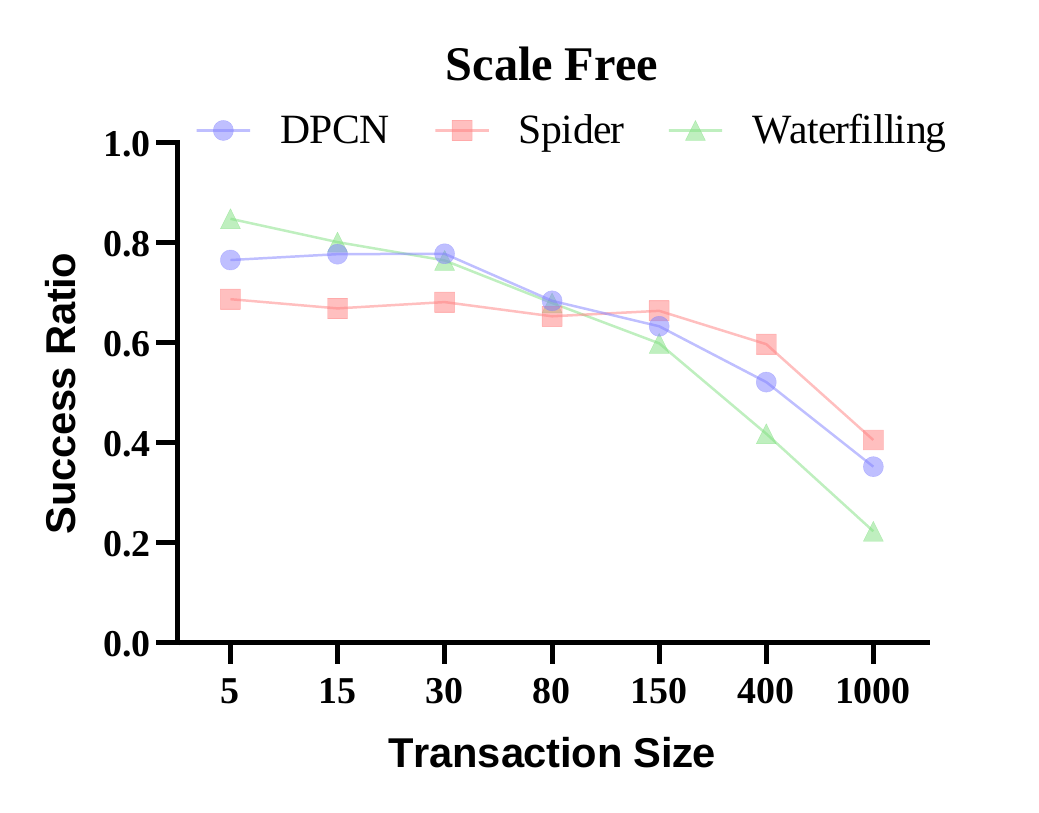}
        \label{sucR4000SF}
    \end{subfigure}
    \hfill
  \begin{subfigure}{0.33\linewidth}
        \includegraphics[width=\linewidth]{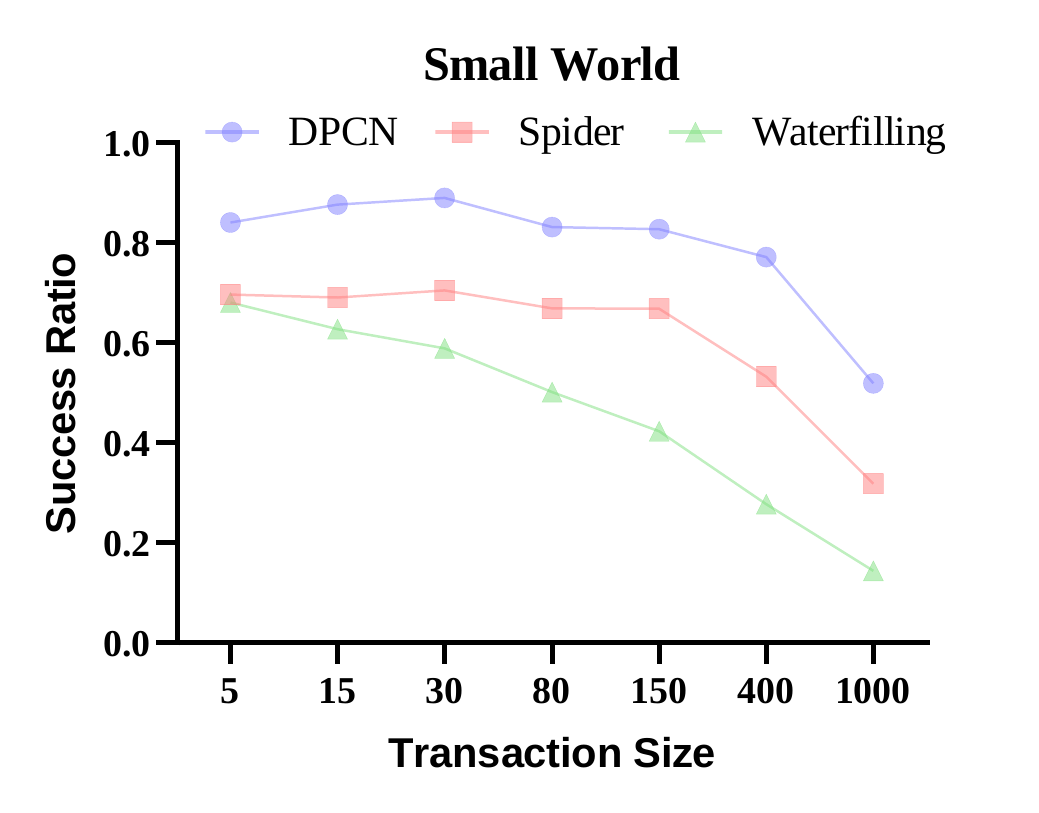}
        \label{sucR4000SW}
    \end{subfigure}
    \hfill
    \begin{subfigure}{0.33\linewidth}
        \includegraphics[width=\linewidth]{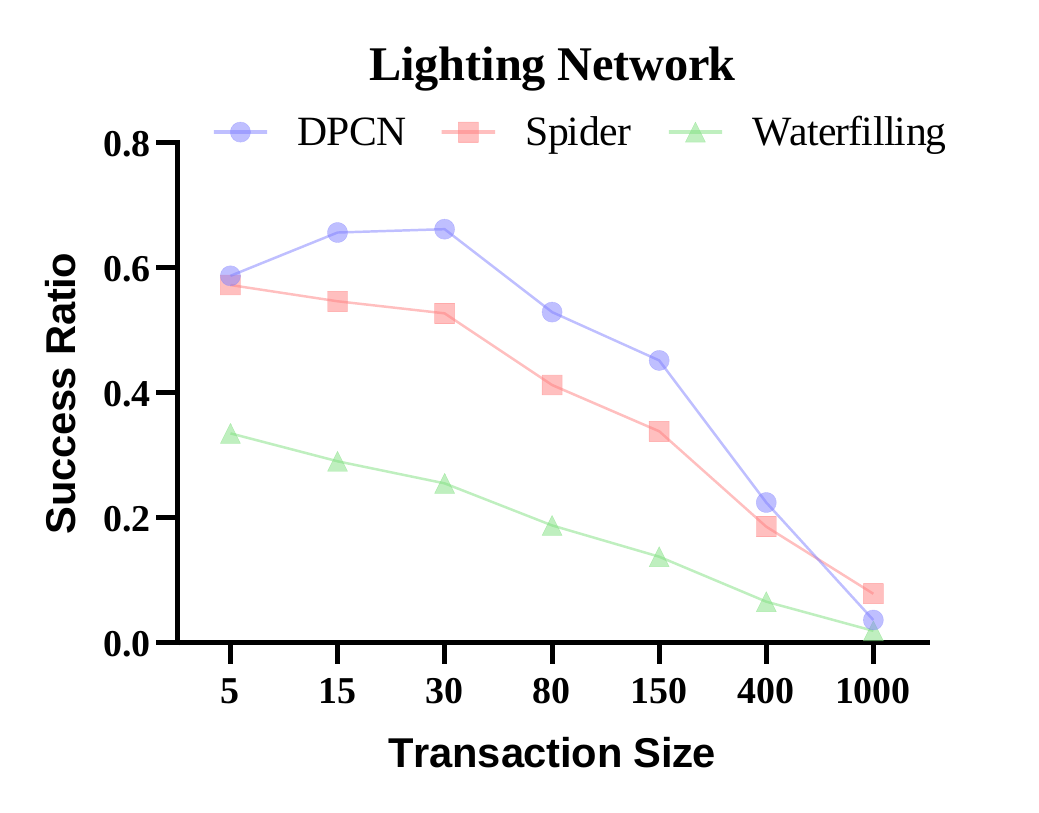}
        \label{sucR4000LND}
    \end{subfigure}
 
    \caption{Success ratio for different transaction sizes of three topologies with mean channel size 4000.}
    \label{sucR4000}
\end{figure*}

\begin{figure*}[ht]
    \centering
   
    \begin{subfigure}{0.33\linewidth}
        \includegraphics[width=\linewidth]{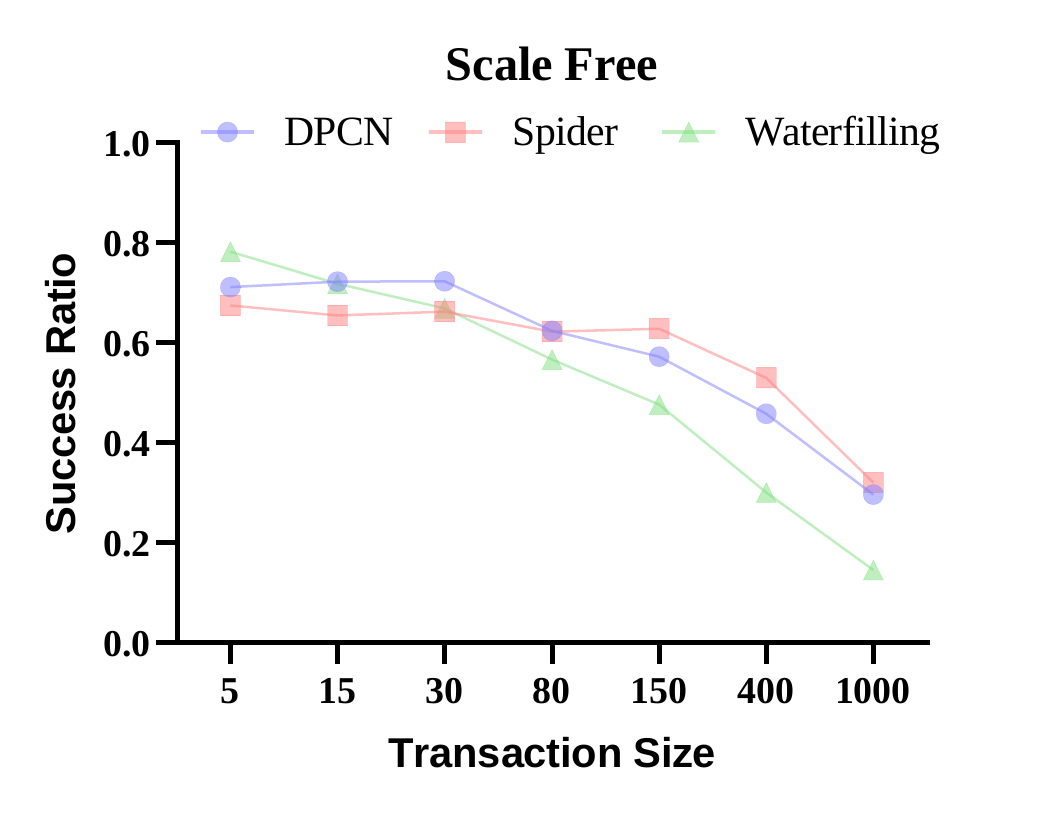}
        \label{sucR2750SF}
    \end{subfigure}    
    \hfill
    \begin{subfigure}{0.33\linewidth}
        \includegraphics[width=\linewidth]{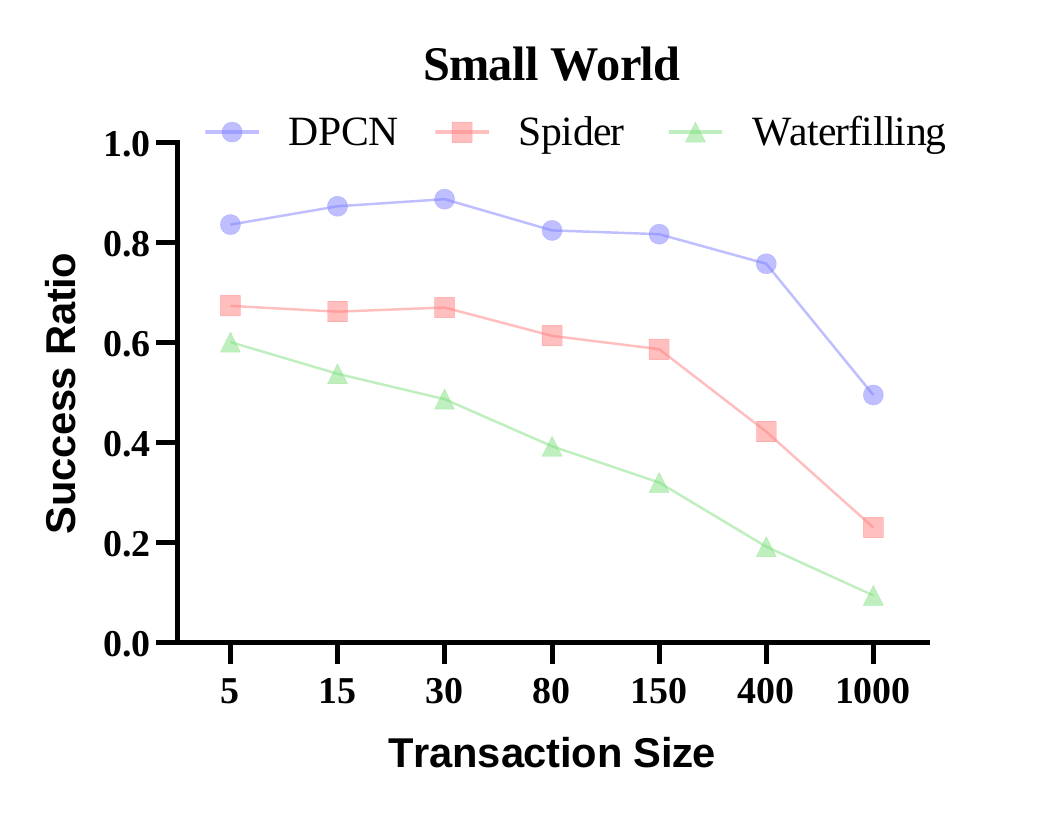}
        \label{sucR2750SW}
    \end{subfigure}
    \hfill
    \begin{subfigure}{0.33\linewidth}
        \includegraphics[width=\linewidth]{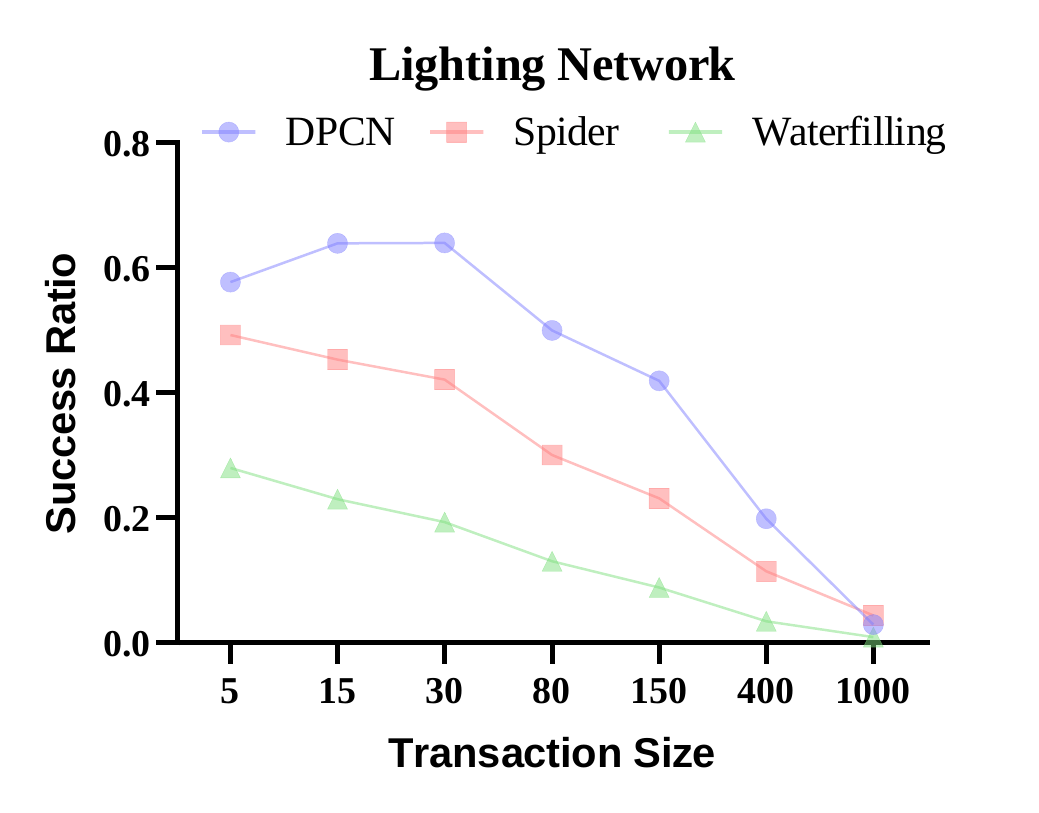}
        \label{sucR2750LND}
    \end{subfigure}
    \caption{Success ratio for different transaction sizes of three topologies with mean channel size 2750}
    \label{sucR2750}
\end{figure*}

\vspace{2mm}
\noindent \textbf{State-of-the-Art Approaches:} We compare our system DPCN with two state-of-the-art approaches:

\begin{itemize}
    \item Spider \cite{spider}: The transaction is sent from the host to the PCNs. Each host has a global topology view of the network and calculates the $k$ edge-disjoint widest paths from the sender to the receiver. For a transaction to be sent, it is first divided into several equal parts according to the maximum transmission unit (MTU) size, and these transactions are sent to these $k$ paths. Only when the window size of a certain path is not less than the sum of the current transaction amount and the unconfirmed transaction amount on the path, the transaction can be processed on the path. Otherwise, the transaction is placed in the end host's queue, and processed according to First-In-First-Out (FIFO) rules. When the transaction queue time at the intermediate routing node exceeds a predetermined threshold, the routing node marks the transaction. After receiving the receiver's acknowledgment message, the router sends this feedback information to the sender with the acknowledgment message, and then the sender adjusts the size of the window of this path according to the mark information.
    
    \item Waterfilling \cite{waterfilling}: Waterfilling also splits the transaction into fund-units and sends these transactions to $k$ edge-disjoint widest paths. For every path, the sender first calculates the minimum channel balance along the path before sending the transaction. After obtaining all paths' balance information, the sender calculates the current available balance of every path based on the bottleneck balance and the size of the unconfirmed transaction, and then sends one split transaction to the path with the largest current available balance. If the maximum available balance of all paths is less than the size of current transaction's fund-units, this transaction is placed in the sender's queue to wait for the next available path.
\end{itemize}

\vspace{2mm}
\noindent \textbf{Metrics:} We use success ratio and success volume as metrics in the evaluation. The success ratio is defined as the ratio of completed transactions to the total number of generated transactions. All the above numbers count the number of original transactions before splitting, i.e. success ratio represents the success rate of original transactions without splitting. The entire transaction is completed only when all split sub-transactions are completed. 
Success volume is equal to the ratio of the total fund-units of successful sub-transactions to the total fund-units of all the sub-transactions. Note that sub-transactions here refer to split transactions.

\vspace{2mm}
\noindent \textbf{Parameters:} For each transaction, we send the split sub-transactions through up to $8$ paths for processing. For the transaction splitting, if a transaction has no deadline, we fix its split size at $20$ fund-units. Otherwise, we set $c$ in equation~\ref{split} to $8$ and $b$ to $0.8$. We set the weighted average $e$ in equation~\ref{eq:alpha} is $0.3$ and $y$ in equation~\ref{estimateTime} is $0.3$. The threshold of waiting time of a transaction in the router $\delta$ mentioned in Section \ref{txCongAvoAlg} is $0.2s$. For scale free and small world networks, we set $g$ in equation~\ref{Wi} as $8.0$ and $\beta$ as $1.3$. For the network with sampled lighting network topology, we set $g$ in equation~\ref{Wi} as $7.0$ and $\beta$ as $0.8$.

\subsection{All Transactions Have Deadlines}
\vspace{2mm}
In the following, we evaluate the performance of different approaches when all transactions have deadlines.

       
      
 

\begin{figure*}[ht]
    \centering
    \begin{subfigure}{0.3\linewidth}
        \fbox{\includegraphics[width=\linewidth]{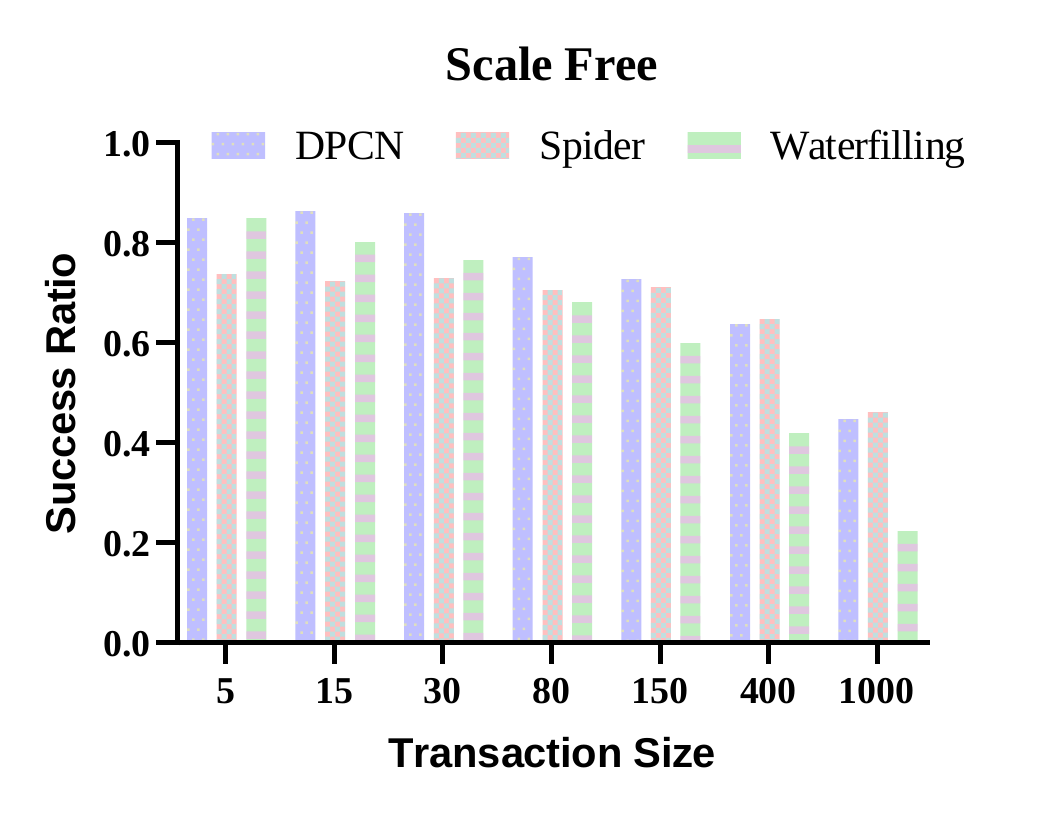}}
        \caption{Mean Channel Size 4000}
        \label{sucR084000SF}
    \end{subfigure}
    \hfill
  \begin{subfigure}{0.3\linewidth}
        \fbox{\includegraphics[width=\linewidth]{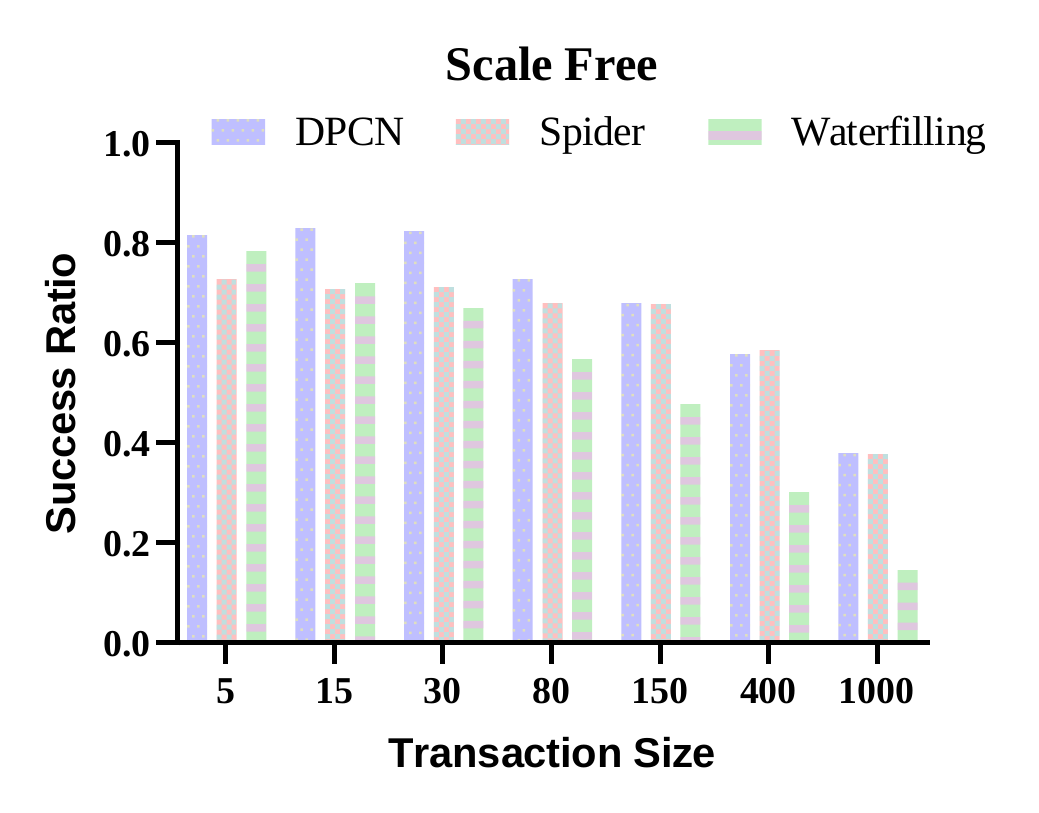}}
        \caption{Mean Channel Size 2750}
        \label{sucR084000SW}
    \end{subfigure}
    \hfill
    \begin{subfigure}{0.3\linewidth}
        \fbox{\includegraphics[width=\linewidth]{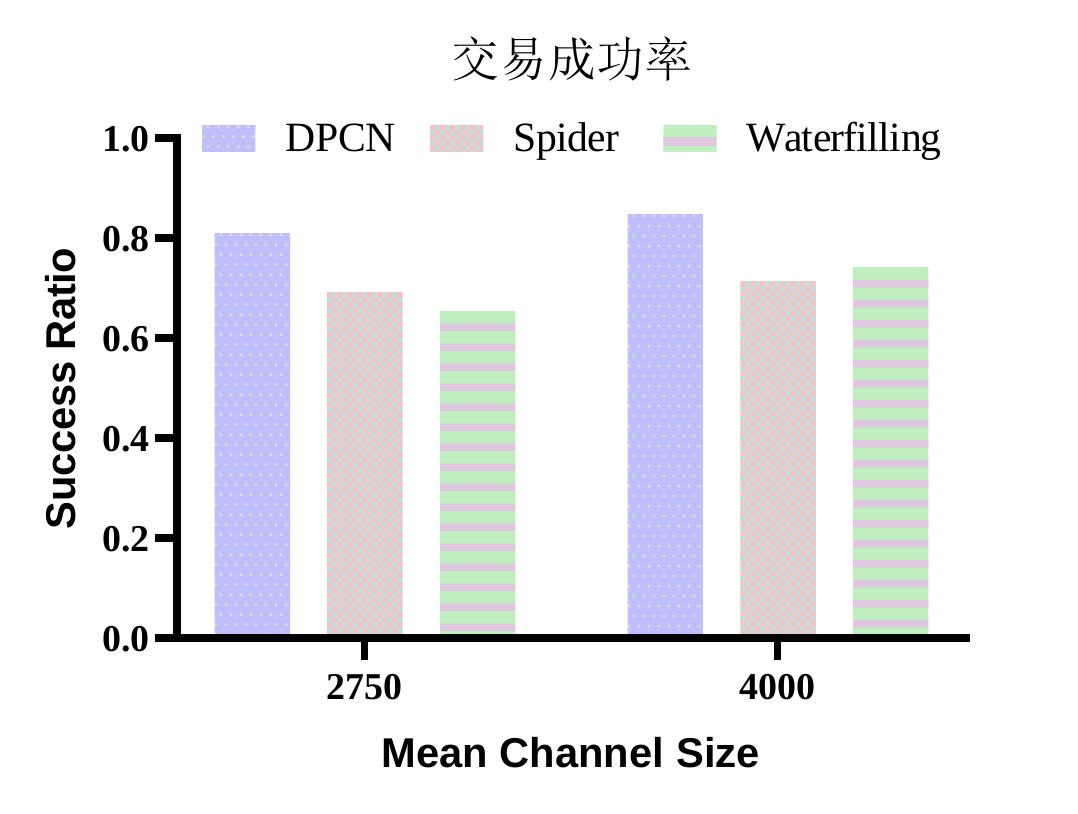}}
        \caption{Success ratio}
        \label{sucR084000ALL}
    \end{subfigure}
    \caption{
    Success ratio for transactions with deadlines}
    \label{sucR08}
\end{figure*}

    
        
       
      

\begin{figure}
    \includegraphics[width=0.8\linewidth]{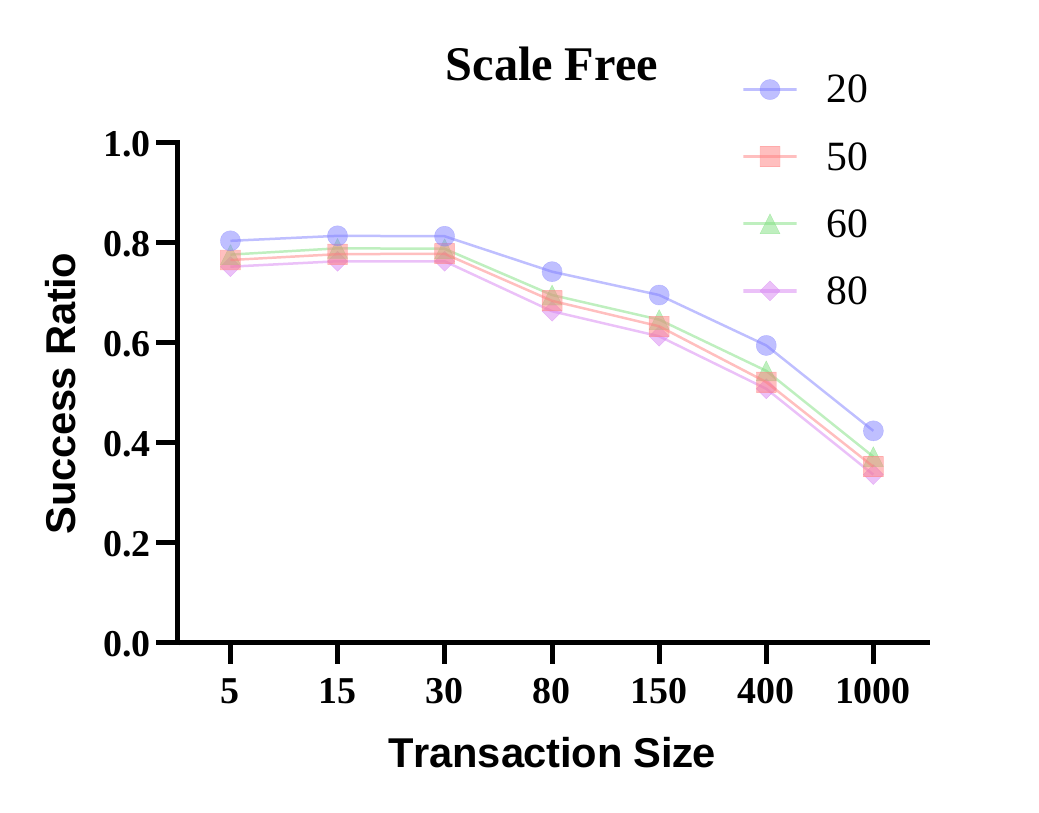}
    \caption{ Mean channel size 4000}
    \label{sf4000DiffN}
\end{figure}

\begin{figure}
    \includegraphics[width=0.8\linewidth]{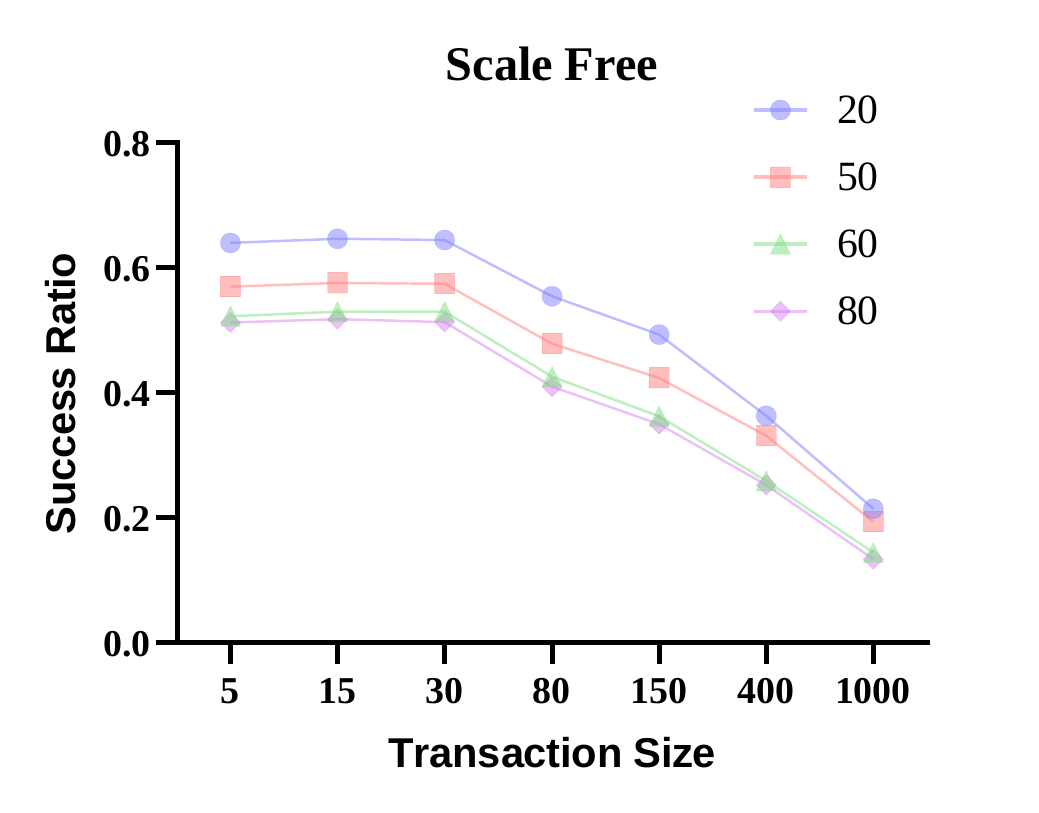}
    \caption{ Mean channel size 900}
    \label{sf900DiffN}
\end{figure}

\vspace{2mm}
\noindent \textbf{Overall performance.} We evaluate the performance of DPCN with other approaches by different mean channel sizes varying from 900 to 8750. As shown in Figure \ref{sucR}, the success ratio of DPCN has a significant improvement compared to the other two approaches. The success ratio is increased by 11\%-40\% compared to Spider, and 26\%-54\% compared to Waterfiling for all three topplogies. Figure \ref{sucV} shows that DPCN also performs better than existing approaches on success volume. The success volume is increased by more than 33\%  compared with Waterfilling and 10\% compared with Spider for most cases. Note that, for the scale-free topology, the result of Spider is better than DPCN in some cases. This is just because the success volume includes the completion of split sub-transactions, Spider splits one transaction into some 1 fund-unit sub-transactions, these transactions are small enough and easier to complete in the network. However, if one of the sub-transactions is not completed because of a timeout, the entire transaction fails. It can be seen from Figures \ref{sucR} and \ref{sucV} that although Spider's success volume is very high, Spider's success ratio is not as high as its success volume. In contrast, our DPCN has a better performance for both success ratio and success volume.

\vspace{2mm}
\noindent \textbf{Performance for different transaction sizes.} Due to space limitation, we only show the success ratio of transactions when the mean channel sizes are 4000 and 2750, respectively. From Figures \ref{sucR4000} and \ref{sucR2750}, we can see that regardless of the size of the mean channel and the transaction, DPCN can effectively improve the success ratio of transactions with deadlines. When the mean channel size is 4000, compared with Spider and Waterfilling, the success ratio increases by more than 10\% in most cases. Compared with Spider, when the transaction amount is between 15 and 80 fund-units, there is an increase of around 15\%. Meanwhile, compared with Waterfilling, for transactions larger than 80 fund-units, the success ratio increases around 22\%. When the mean channel size is 2750, the success ratio increases 19\% and 25\%, respectively.

\begin{figure}
    \includegraphics[width=0.8\linewidth]{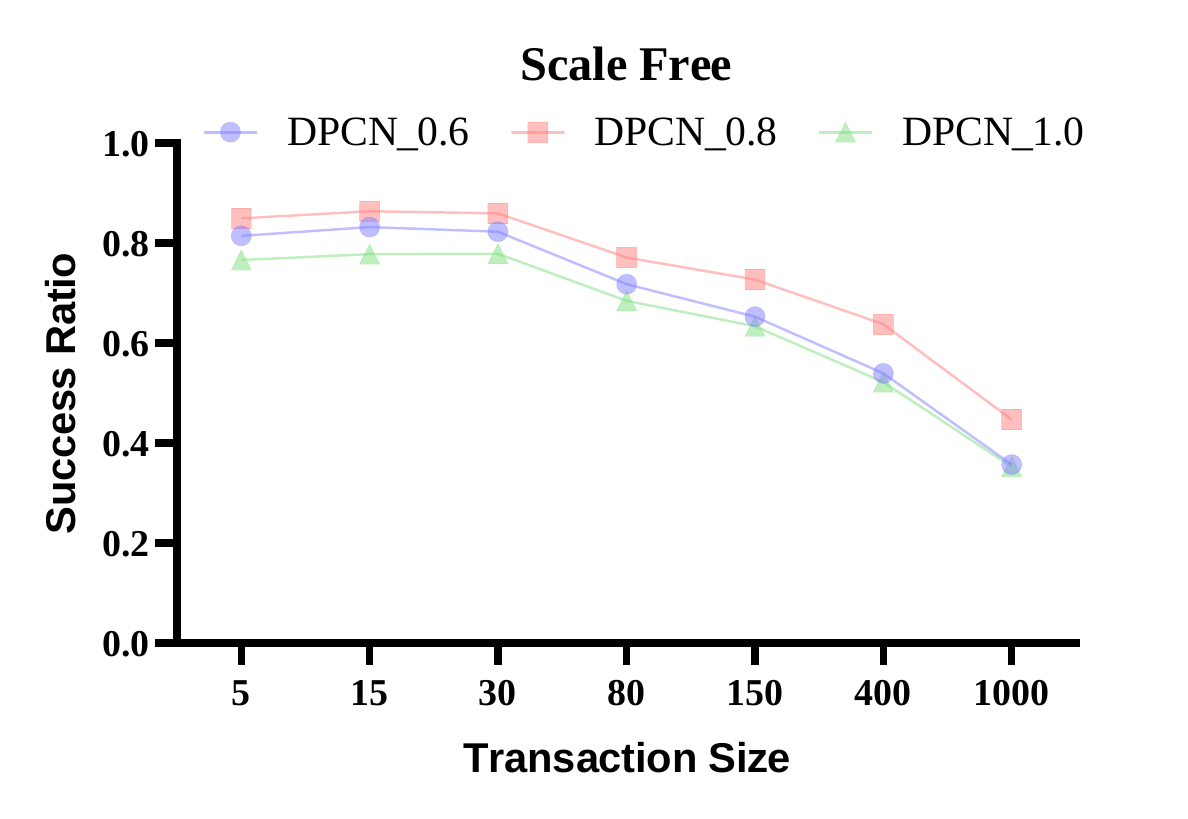}
    \caption{Mean channel size 4000}
    \label{sfDiffR4000}
\end{figure}

 \begin{figure}
    \includegraphics[width=0.8\linewidth]{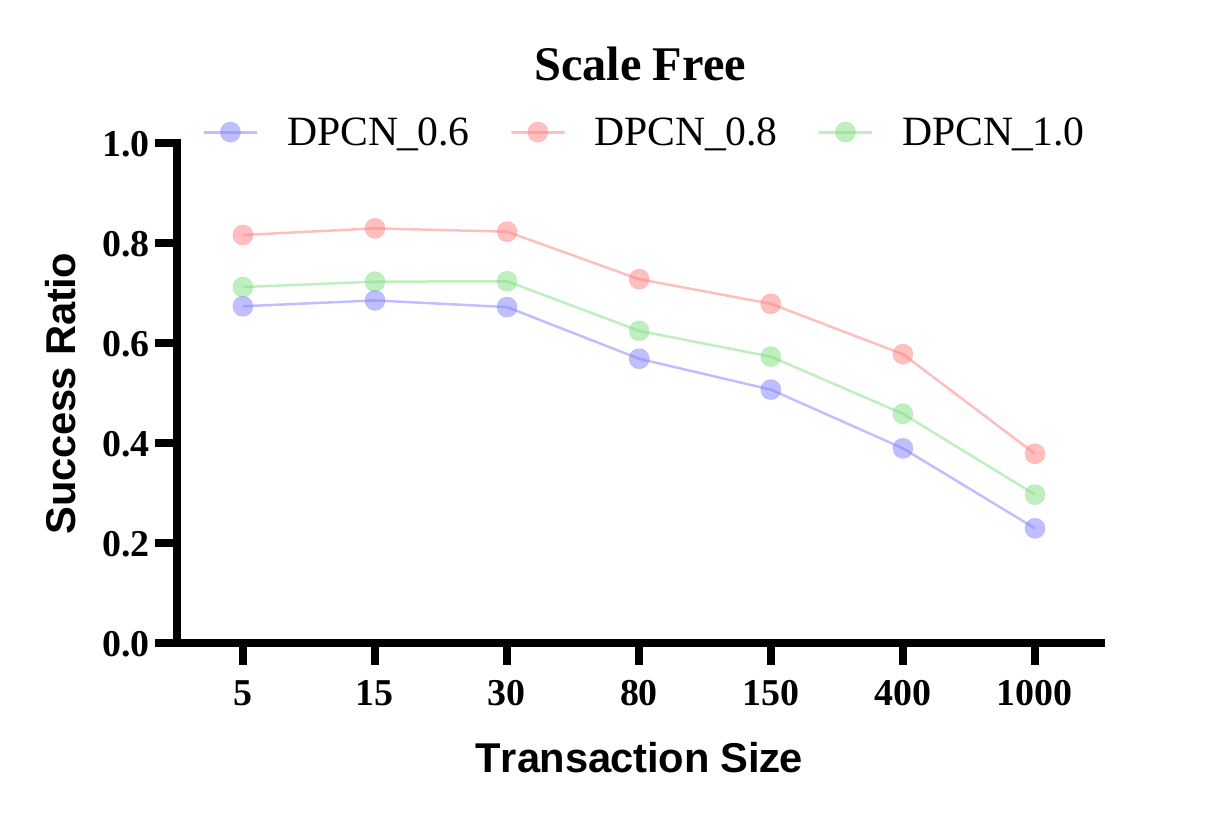}
    \caption{Mean channel size 2750}
    \label{sfDiffR2750}
\end{figure}

\vspace{2mm}
\noindent \textbf{Performance for different network sizes.} We use the scale-free topology and set the mean channel size to 900 and 4000. We increase the number of nodes in the network from 20 to 80, and compare success ratios as shown in Figures \ref{sf4000DiffN} and \ref{sf900DiffN}.
When the  network size is small, the number of channels required to complete each transaction is smaller, and the processing time is shorter, so the success ratio is higher at this time. However, when the number of nodes in the network is too large, the topology of the network will become more complex, and the path required for a transaction to complete may become longer (the number of channels needs to be increased), which will increase the processing time of a transaction. Therefore, when the number of nodes in the network increases from 20 to 80, the success ratio of transactions in the network decreases.
However, from Figures \ref{sf4000DiffN} and \ref{sf900DiffN}, we can see that the change in the number of nodes has a minor impact on the success ratio, which indicates that our DPCN is robust enough to handle networks with different nodes.

\subsection{Mixing Deadline and Non-deadline Transactions}

\vspace{2mm}
\noindent \textbf{Performance for different transaction sizes.} We evaluate our DPCN when transactions with deadlines and without deadlines are mixed. As shown in Table \ref{TxDDL}, we have seven transaction sizes in our evaluation. We randomly select 20\% of transactions with each size and set these transactions to have no deadline. Since the arrival order of transactions is random, these transactions with no deadline may arrive earlier than those with deadlines and have already occupied network resources for processing, resulting in that transactions with deadlines cannot be processed in a timely manner. 
As shown in Figures\ref{sucR08} , for scale-free topology, DPCN has an average of 5\% and 14\% increase in the success ratio of transactions with deadline, compared to Spider and Waterfilling respectively. This is because existing approaches are deadline agnostic, and transactions without deadlines are likely to occupy the resources of time-critical transactions. Different from these approaches, DPCN (1) prioritizes transactions with tight deadlines as much as possible, (2)makes the size of time-critical transactions as large as possible, thereby reducing transaction processing time, and (3) dynamically adjusts the window size according to the deadline and the completion time of the transactions. Even under the interference of transactions without deadlines, transactions with deadlines still have a higher success ratio.

\vspace{2mm}
\noindent \textbf{Performance for different ratios of transactions with deadlines.} We evaluate DPCN's performance when the proportion of transactions with deadlines is different. We set the topology structure to scale-free and the proportion varies from 0.6 to 1.0. We compare the success ratio when the mean channel sizes are 4000 and 2750 respectively. From Figure \ref{sfDiffR4000} and  \ref{sfDiffR2750}, we can see that our DPCN is robust enough to handle a wide spectrum of deadline distributions and can effectively improve the success ratio of transactions with deadlines.

\section{Related Work}
\label{related}
\vspace{2mm}

\vspace{2mm}
The first payment channel networks proposed by the Lightning Network\cite{lightingnetwork} routed transactions along the shortest path. Although transactions on Lighting Network can be completed in a shorter time than processing transactions on blockchain, the Lightning Network still faces the problem of insufficient transaction success ratio and throughput. In order to solve these problems, various algorithms have been proposed for PCNs. Flash\cite{flash} differentiated transactions of different sizes. For small transactions, they used the routing table to route the transaction based on the shortest path; for large transactions, they used the modified maximum flow algorithm for transaction routing. CoinExpress\cite{coinexpress} modified the Ford-Fulkerson algorithm for maximum flow\cite{maximal}, and proposed a distributed approach to route the transaction. These approaches were based on network flow. In addition, there were approaches based on landmark nodes\cite{landmark} and approaches based on data networks. SilentWhispers\cite{silentwhispers} proposed a two-way breadth-first search(BFS) to find the shortest path from the sender to the landmark and the shortest path from the landmark to the receiver. They used this path through the landmark as the routing path of the transaction. In order to solve privacy disclosure problem such as balances in the payment channel of intermediate nodes, this paper proposes a secret-sharing-based multi-party computation and digital signature chains to protect private data in the payment channel networks. Different from SilentWhispers, the landmark node in SpeedyMurmurs\cite{SpeedyMurmurs} used the tree-based graph embedding approach to provide the transaction path. Compared to approaches based on network flow and landmark nodes, the approaches based on data networks fully considered dynamics of the actual payment channel networks. In Spider \cite{spider}, similar to data packets in a data network, each transaction is split into several fund-units and routed through different paths. \cite{scheDeadline} performed a good study on the single-hop optimal transaction scheduling problem with the deadline consideration. Other remotely related work is as follows. D3\cite{D3} required the sender, receiver, and switch to participate in the optimization, the core idea of which is to implement differentiated services for various streams based on deadlines. D2TCP\cite{d2tcp} combined the advantages of DCTCP and D3. It considered not only the congestion avoidance, but also the flow deadline. 
However, these approaches, which inspired us in our design, cannot be simply applied to the PCNs, as the problem we solve in PCNs differs from it in crucial ways. In the Internet, the capacity of each link is unchanged when sending packets, while the capacity of each channel varies dynamically after every payment in the channel. PCNs are more dynamic than Internet and it is more challenging to guarantee transactions' completion time in PCNs.

\section{CONCLUSION}
\label{con}
\vspace{2mm}
In this paper, we have presented a new and systematic framework DPCN, which considers transaction deadlines in the payment channel networks. We have proposed to enable such framework via a synergy of three techniques, deadline-based dynamic transaction split, deadline-aware transaction scheduling and congestion avoidance algorithm. Our framework DPCN can ensure that time-sensitive transactions can be completed within the specified time, while improving the success ratio of transactions of PCNs. Our extensive evaluation results have shown that DPCN has significantly outperformed the state-of-the-art approaches.

\ifCLASSOPTIONcaptionsoff
  \newpage
\fi



\bibliographystyle{IEEEtran}
\bibliography{IEEEabrv,example}
\end{document}